\documentclass[11pt]{article}
\usepackage{jcappub}
\usepackage{mathtools}
\usepackage{bm}
\usepackage{dsfont}
\usepackage{array}
\usepackage{graphicx}
\usepackage{subcaption}
\usepackage{soul}
\usepackage{multirow}
\usepackage{multicol}
\usepackage{float}
\usepackage{hhline}
\usepackage[dvipsnames]{xcolor}
\usepackage[normalem]{ulem}
\usepackage{mathrsfs}
\usepackage{wasysym}
\usepackage[mathscr]{euscript}

\title{Stars Within Stars: The Interplay Between Main-Sequence Stars and Dark Matter}

\author[a]{Boris Betancourt Kamenetskaia,}
\author[b]{Chris Kouvaris}

\affiliation[a]{Cosmology, Gravity, and Astroparticle Physics Group, Center for Theoretical Physics of the Universe,
Institute for Basic Science (IBS), Daejeon, 34126, Korea}
\affiliation[b]{Physics Division, National Technical University of Athens, 15780 Zografou Campus, Athens, Greece}

\emailAdd{laybors@ibs.re.kr}
\emailAdd{kouvaris@mail.ntua.gr}

\abstract{Asymmetric dark matter may form self-gravitating macroscopic structures called dark stars if its microscopic properties give rise to sufficient effective pressure to counteract gravity. We study low-mass baryonic stars embedded within such dark stars, treating the baryonic and dark components as two gravitating fluids. The dark sector is assumed to be non-luminous and to interact with the baryonic star only through gravity, while the baryonic structure is computed using a realistic equation of state, opacity tables, and proton-proton-chain energy production. We find that the additional gravitational potential increases the luminosity at fixed baryonic mass, as the baryons must reach higher temperatures to maintain hydrostatic equilibrium. This lowers the minimum baryonic mass for hydrogen-burning configurations below the standard value of $\sim0.085M_\odot$. The resulting increase in luminosity can reduce the estimated main-sequence lifetime significantly, bringing low-mass configurations that would otherwise remain on the main sequence for hundreds of Gyr to cosmologically relevant timescales. We further show that photons escaping from the stellar surface undergo gravitational redshift determined by the global dark matter mass distribution. Consequently, extended dark star configurations can produce distinct observational signatures even when their outer regions have little effect on the intrinsic baryonic stellar structure, providing a potential probe of otherwise hidden dark structures.}

\begin{document}
\maketitle
\flushbottom

\section{Introduction}
Dark matter (DM) constitutes approximately $27\%$ of the energy density of the Universe~\cite{Planck:2018vyg}, yet its nature remains one of the central open questions in modern physics. Although its existence is firmly established through observations ranging from galactic rotation curves to cosmological measurements, it remains unknown whether DM consists of elementary particles, compact astrophysical objects, or some entirely different form of matter (for reviews on different DM models, see \cite{Bertone:2004pz,Feng:2010gw, SciPostPhysRev.1}). Among the various possibilities, particle DM models have been extensively studied  due to both strong theoretical motivation and ability to account for the observed cosmological and astrophysical properties of DM.

Among the various particle DM theories there is a strongly motivated class of models
which is that of asymmetric DM (ADM)~\cite{Nussinov:1985xr,Barr:1990ca,Barr:1991qn,Gudnason:2006ug,Kouvaris:2007iq}. ADM assumes a conserved quantum number and an asymmetry in the populations of particles and antiparticles. In early times DM annihilations deplete the population of antiparticles leaving only a substantial population of particles that (as expected) can no longer annihilate. ADM models are well theoretically motivated because they can provide a link between the baryonic and DM sectors since the asymmetry in both sectors might be related. 
A particularly fascinating consequence of the ADM paradigm is the possibility that DM can form gravitationally bound, macroscopic configurations if its microscopic properties give rise to an effective pressure capable of counteracting gravity, provided that a mechanism for energy evacuation is in place in order to facilitate the collapse and the subsequent formation. The pressure can originate from quantum effects, self‑interactions, or a combination of both, depending on the underlying particle model. This idea dates back to Kaup~\cite{Kaup:1968zz}, who studied a self‑gravitating system of complex scalar fields, and has since been extended to a wide variety of particle models: bosonic DM with repulsive or attractive self‑interactions \cite{Colpi:1986ye,Eby:2015hsq}, fermionic DM supported by degeneracy pressure or self‑interactions \cite{Kouvaris:2015rea}, and many others. Collectively, these objects are often referred to as dark stars (DSs). Regardless of the underlying particle physics, they are described by hydrostatic equilibrium, with their masses, radii, and internal density profiles determined by the DM equation of state (EoS). Depending on the DM particle mass and interaction strength, DSs can span many orders of magnitude in size and mass, ranging from subsolar masses to above $10^{5}\,M_\odot$, with radii extending from compact stellar scales to several solar radii or more, while remaining stable over cosmological timescales.

Several observational techniques constrain the presence of compact dark objects through their gravitational interactions. These include, for example, stellar microlensing~\cite{Macho:2000nvd,EROS-2:2006ryy,Niikura:2019kqi}, supernova magnification~\cite{Zumalacarregui:2017qqd}, gravitational wave searches for sub‑solar‑mass binaries~\cite{LVK:2022ydq}, and the dynamical heating of dwarf galaxies~\cite{Brandt:2016aco}. Although originally developed to search for compact DM candidates such as Massive Compact Halo Objects (MACHOs) and primordial black holes (PBHs), these observations can also be interpreted as limits on compact DSs, implying that they can constitute at most a small fraction of the total DM population, namely a few percent~\cite{Green:2020jor,SciPostPhysRev.1}. If such structures exist, ordinary baryonic stars may form inside them, be captured by them, or otherwise become embedded in their gravitational potential. In this situation, the dark component can substantially modify the baryonic star without necessarily providing an additional source of stellar energy.

The interplay between DM and stellar objects has been explored extensively in the context of the first stars. The possibility that DM annihilation could provide an important source of energy during primordial star formation led to the proposal of annihilation-powered dark stars, baryonic stellar objects in which DM heating can temporarily dominate over nuclear fusion~\cite{Spolyar:2007qv}. Detailed studies subsequently demonstrated that such objects can have stellar structures markedly different from those of conventional Population III stars, including larger radii, lower surface temperatures, and potentially very large masses~\cite{Freese:2008wh,2009ApJ...705.1031S,Lopes:2014xaa}. These annihilation‑powered objects should not be confused with the dark objects studied in this work. Here, the term dark star (DS) refers to a self‑gravitating configuration made entirely of ADM and therefore no DM annihilations are present.

A related class of studies has considered the capture of DM particles by ordinary stars. In that case, scattering between DM and baryons can replenish a population of DM particles gravitationally bound to the star, allowing continuous DM annihilation that can keep an old neutron star warmer than expected~\cite{Kouvaris:2007ay,Baryakhtar:2017dbj} and/or extend a DM‑powered stellar phase~\cite{Freese:2008wh}. Captured DM can also transport energy within the star and thereby modify the temperature gradient and the conditions for nuclear burning~\cite{Iocco:2008rb,Taoso:2010tg} (see for example Ref.~\cite{Lopes:2011rx} for detailed calculations of the capture rate through the evolution of low-mass stars, including how stellar structure feedback affects the capture process itself). Most of this literature, however, assumes that DM is a trace component inside the baryonic star and that its influence arises through annihilation, energy injection, or energy transport.

Another closely related body of work has studied admixed stars, in which a DM component coexists with ordinary baryonic matter. Most of these studies focus on neutron stars containing a DM core or admixture, described within a two-fluid framework. Even a small DM fraction can substantially modify the stellar structure, including the mass–radius relation \cite{Leung:2011zz,Xiang:2013xwa}, tidal deformability \cite{Ellis:2018bkr,Giangrandi:2022wht}, and stability \cite{Leung:2012vea}, and can lead to equilibrium configurations with no standard-neutron-star analogue \cite{Leung:2011zz,Leung:2012vea}. Such DM-admixed neutron stars may leave observable imprints in gravitational-wave signals; in particular, mergers involving DM cores could be distinguishable from ordinary neutron-star mergers \cite{Ellis:2017jgp,Emma:2022xjs}. A broad range of hybrid admixed compact-star models has since been explored \cite{Karkevandi:2021ygv,Kain:2021hpk,Deliyergiyev:2019vti,Leung:2022wcf,Das:2020ecp,Routaray:2022utr}, including effects on stellar structure and cooling \cite{Leung:2011zz,Leung:2012vea,Kain:2021hpk,Miao:2022rqj,Ivanytskyi:2019wxd,Karkevandi:2021ygv,Leung:2022wcf}.

In this paper we consider a complementary situation: a low‑mass baryonic star embedded in a substantial, extended, and self‑gravitating dark component. We assume that the dark component does not provide an additional source of stellar energy and instead affects the baryonic object only through its gravitational field. The baryonic star and the dark component are treated as two gravitating fluids with separate equations of state. The baryonic component obeys the usual equations of stellar structure, while the enclosed dark mass contributes to the hydrostatic support of the baryons. This setup is particularly interesting for low‑mass stars, whose central temperatures are close to the hydrogen‑burning threshold and whose structure is therefore especially sensitive to changes in the gravitational potential.

The purpose of this work is to establish a framework for studying stars embedded in self‑gravitating dark environments and to quantify how the properties of the dark component are imprinted on the baryonic star. Rather than considering DM as a source of stellar heating, we isolate its gravitational effect and study how different dark density profiles and EoS modify stellar structure, the minimum mass for hydrogen burning, stellar luminosities, evolutionary timescales, and observable gravitational redshifts.

\section{Equations of stellar structure}
\subsection{Standard scenario}
We will firstly present for pedagogical purposes a simplified version of the equations that describe a main sequence star without the presence of DM. The equations of stellar structure are given by
\begin{align}
    \frac{dm}{dr} &= 4\pi r^2 \rho, \label{eq:mass_cont} \\
    \frac{dP}{dr} &= -\frac{Gm\rho}{r^2}, \label{eq:hydro_eq} \\
    \frac{d\ell}{dr} &= 4\pi r^2 \rho \epsilon, \label{eq:energy_prod} \\
    \frac{dT}{dr} &= -\nabla \frac{T}{P}\frac{Gm\rho}{r^2}. \label{eq:energy_transport}
\end{align}
Equation (\ref{eq:mass_cont}) is the mass continuity equation with $m$ the enclosed mass at radius $r$, (\ref{eq:hydro_eq}) is the hydrostatic equilibrium equation, (\ref{eq:energy_prod}) is the energy production equation with $\ell$ the luminosity at radius $r$ and (\ref{eq:energy_transport}) is the energy transport equation. These four equations, augmented by the equation of state (EoS), the energy generation rate per unit mass $\epsilon$ and the opacity $\kappa$, form the system of equations of stellar structure, which once solved yield the mass, pressure, luminosity and temperature profiles of the star. The $\nabla$ is defined later in the text below.

Regarding the EoS, we model a star with constant mass fractions $X_1=X_{\rm H}=X$, $X_2=X_{\rm He}=Y$ and $X_3=X_{\rm heavy}=Z$ such that $X+Y+Z=1$. For orientation, the pressure equation of state of a main-sequence star can be approximated by
\begin{equation}\label{eq:P_EoS}
    P(\rho,T,X,Z)=P_{\rm deg}+P_{\rm gas}+P_{\rm rad}=\frac{1}{5}\left(\frac{3}{8\pi}\right)^{\frac23}\frac{h^2}{m_e m_u^{\frac53}\mu_e^{\frac53}}\rho^{\frac{5}{3}}+\frac{\rho k_B T}{\mu m_u}+\frac{1}{3}\alpha T^4,
\end{equation}
where $m_u=0.93\ \mathrm{GeV}/c^2$ is the atomic mass unit, $\alpha=\frac{4}{c}\sigma_{\rm SB}$ is the radiative constant with $\sigma_{\rm SB}=\frac{\pi^2k_B^4}{60c^2\hbar^3}$ the Stefan-Boltzmann constant. The term $\mu$ is the mean molecular weight defined as
\begin{equation}\label{eq:mu}
    \mu=\left[\sum_{j}{\frac{(Z_j+1)X_j}{A_j}}\right]^{-1}\approx\frac{4}{3+5X-Z},
\end{equation}
where the sum goes over all elements in the star with $A_j$ and $Z_j$ the mass and atomic number of element $j$, respectively. Within this approximation, we have assumed  $\frac{Z_j+1}{A_j}\approx\frac12$ for heavy elements and have written $Y=1-X-Z$. Additionally, the expression $\mu_e$ is the mean molecular weight per electron, such that
\begin{equation}
    \mu_e=\frac{\rho}{m_u n_e}\approx\frac{2}{1+X},
\end{equation}
where $n_e$ is the electron number density. We have used the same approximations as in eq.~\eqref{eq:mu} for the expression on the right-hand side. These analytic expressions are useful for illustrating the leading contributions to the pressure and the composition dependence of the baryonic matter. In the actual stellar-structure calculations, however, we adopt the more complete, tabulated equation of state from Ref.~\cite{2019ApJ...872...51C}, evaluated for the same composition variables $X$ and $Z$.\footnote{EoS table downloaded from \url{http://perso.ens-lyon.fr/gilles.chabrier/DirEOS}.}

For the energy generation, the total nuclear energy generation rate per unit mass $\varepsilon$ (in units of $\mathrm{W}\,\mathrm{kg}^{-1}$) due to hydrogen burning via the proton–proton chain is computed as
\begin{align}\label{eq:energy_generation}
\varepsilon &=
r_{\mathrm{pp}}(Q_{\mathrm{pp}}+Q_{\mathrm{Dp}})\cr
&+r_{^3{\rm He}^3{\rm He}}Q_{^3{\rm He}^3{\rm He}}
+r_{^3{\rm He}^4{\rm He}}Q_{^3{\rm He}^4{\rm He}}
+r_{^7{\rm Be}\,{\rm e}^-}Q_{^7{\rm Be}\,{\rm e}^-}
+r_{^7{\rm Li}\,{\rm p}}Q_{^7{\rm Li}\,{\rm p}}
+r_{^7{\rm Be}\,{\rm p}}Q_{^7{\rm Be}\,{\rm p}},
\end{align}
where $Q_i$ is the energy released per reaction (given in Appendix~\ref{app:energy_injection}) and the per‑mass reaction rates $r_i$ (in units of $\mathrm{s}^{-1}\,\mathrm{kg}^{-1}$) are derived from temperature‑dependent cross sections $N_{A}\langle\sigma v\rangle_{i}(T)$ (taken from the NACRE compilation~\cite{Angulo:1999zz}). The explicit expressions for the rates $r_i$ in terms of the number densities, the steady‑state adjustments for the intermediate nuclei ($^3\mathrm{He}$, $^7\mathrm{Be}$, $^7\mathrm{Li}$), and the full forms of $N_{A}\langle\sigma v\rangle_{i}(T)$ are all provided in Appendix~\ref{app:energy_injection}.

Finally, concerning the opacity $\kappa$, we consider a fixed composition of $X=0.7$ and $Z=0.02$. For small temperatures ($2<\log_{10}(T/{\rm K})<4.5$) we use the table obtained with the {\AE}SOPUS code~\cite{2009A&A...508.1539M,2022ApJ...940..129M,2024ApJ...976...39M} for the chemical composition of Magg et al.~\cite{Magg:2022rxb}\footnote{{\AE}SOPUS opacity tables are provided in \url{https://stev.oapd.inaf.it/aesopus_2.1/tables/tables/SOLAR/}.}. At higher temperatures ($4.5<\log_{10}(T/{\rm K})<9.05$), we adopt the tables obtained with the OPAL code~\cite{1996ApJ...464..943I} for the composition of Grevesse and Noels~\cite{1993oee..conf.....P}\footnote{OPAL opacity tables are provided in \url{https://opalopacity.llnl.gov/existing.html}.}.

Assuming that energy can only be transported radiatively or convectively, we have \citep{2004sipp.book.....H}
\begin{equation}
    \nabla=
    \begin{cases}
        \nabla_{\rm rad},\hspace{0.5cm}\text{if }\nabla_{\rm rad}\leq\nabla_{\rm ad}\ \text{(diffusive radiative transfer)}\cr
        \nabla_{\rm ad},\hspace{0.6cm}\text{if }\nabla_{\rm rad}>\nabla_{\rm ad}\ \text{(adiabatic convection)}.
    \end{cases}
\end{equation}
Here, $\nabla$ is the local slope of temperature with respect to pressure, given by
\begin{equation}
    \nabla=\frac{\partial \ln T}{\partial \ln P},
\end{equation}
and for the radiative process we have
\begin{equation}
    \nabla_{\rm rad}=\frac{3}{16\pi \alpha c}\frac{P\kappa}{T^4}\frac{\ell}{Gm}.
\end{equation}
In the case of convection and an EoS given by eq.~\eqref{eq:P_EoS}, we have
\begin{equation}
    \nabla_{\rm ad}=\frac{\Gamma_2-1}{\Gamma_2}=\frac{2(4-3\beta)}{40-35\beta-(3\beta^2-11\beta+8)\beta^\prime},
\end{equation}
where $\Gamma_2$ is the second adiabatic exponent with $\beta\equiv P_{\rm gas}/(P_{\rm gas}+P_{\rm rad})$ and $\beta^\prime\equiv (P_{\rm gas}+P_{\rm rad})/(P_{\rm deg}+P_{\rm gas}+P_{\rm rad})$.

\paragraph{Boundary conditions.} The equations of stellar structure must be supplemented by suitable boundary conditions. Formally, we have at the center of the star ($r=0$):
\begin{equation}
    m(0)=\ell(0)=0,\, P(0)=P_c=P(\rho_c,T_c,X,Z),\, T(0)=T_c,
\end{equation}
where $\rho_c$ and $T_c$ are the core density and temperature, respectively. The core pressure $P_c$ is then fixed by the equation of state (\ref{eq:P_EoS}). Nevertheless, it proves useful for numerical purposes to set boundary conditions at a slightly larger radius. For small $r_\delta$ (i.e., near the center), we have the approximate behavior:
\begin{align}
    m(r_\delta) &\approx \frac{4\pi}{3}r_\delta^3\rho_c,\cr
    P(r_\delta) &\approx P_c-\frac{2\pi G}{3}\rho_c^2r_\delta^2,\cr
    \ell(r_\delta) &\approx \frac{4\pi}{3}r_\delta^3\rho_c\epsilon_c,\cr
    T(r_\delta) &\approx
    \begin{cases}
        T_c\,\left[1-\frac{\kappa_c\rho_c^2\epsilon_c}{2acT_c^4}r_\delta^2\right]^{\frac{1}{4}},\hspace{0.5cm}\text{if }\nabla_{\rm rad,c}\leq\nabla_{\rm ad,c}\cr
        T_c\,\mathrm{e}^{-\frac{2\pi G}{3}\frac{\rho_c^2}{P_c}\nabla_{\rm ad,c}r_\delta^2},\hspace{0.8cm}\text{if }\nabla_{\rm ad,c}\leq\nabla_{\rm rad,c}
    \end{cases},
\end{align}
where the subscript $c$ stands for the given quantity evaluated at the core of the star. However, not all choices of $\rho_c$ and $T_c$ lead to a valid model. The additional constraints required to select a unique solution are provided by the boundary conditions at the stellar surface ($r=R$).

Formally, the surface of a star is defined as the point where the pressure and density vanish ($P(R)=\rho(R)=0$). In practice, this is not directly usable in a numerical integration, as the equations of stellar structure become singular when $\rho$ and $P$ approach zero. Additionally, our physical description of the stellar interior, namely the equation of state, opacities, and energy generation, is not valid in the low-density, cool outer layers. Therefore, we employ a more practical surface condition. We define the stellar radius $R$ as the point where the optical depth $\tau$ equals $2/3$. This is the location of the photosphere, from which most of the radiation escapes and which corresponds to the visible ``surface'' of the star. In the context of the stellar structure equations, this condition is evaluated using the hydrostatic equilibrium approximation for the atmosphere:
\begin{equation}
\tau(R) \approx \int_R^{\infty} \kappa \rho dr = \frac{2}{3}.
\end{equation}
This integral can be approximated using the local pressure and gravity, leading to the common photospheric boundary condition:
\begin{equation}
\label{eq:tau_boundary}
\tau \approx \frac{\kappa P}{g} = \frac{2}{3}, \quad \text{where} \quad g = \frac{G m(R)}{R^{2}}.
\end{equation}
The integration is stopped when this condition is met. At this point, we identify the radius $R$, the total mass $M=m(R)$, and the luminosity $L=\ell(R)$. The final boundary condition links the luminosity to the temperature at this photospheric radius. Assuming the star radiates as a black body, the luminosity is given by the Stefan–Boltzmann law:
\begin{equation}
\label{eq:luminosity_boundary}
L = 4\pi R^{2} \sigma_{\mathrm{SB}} T_{\mathrm{eff}}^{4},
\end{equation}
where $\sigma_{\mathrm{SB}}$ is the Stefan–Boltzmann constant and $T_{\mathrm{eff}}=T(R)$ is the effective temperature.

These two surface conditions are not independent; together they select a unique stellar model for a given central temperature $T_c$ and composition. In practice, for a fixed $T_c$, the central density $\rho_c$ is treated as a free parameter. The structure equations are integrated outward until the photospheric condition $\tau = 2/3$ is reached, yielding a trial radius $R$, luminosity $L$, and effective temperature $T_{\mathrm{eff}}$. The central density is then adjusted iteratively until the Stefan–Boltzmann relation $L = 4\pi R^{2} \sigma_{\mathrm{SB}} T_{\mathrm{eff}}^{4}$ is satisfied. Hence, each choice of $T_c$ yields a consistent stellar model.
\begin{figure}[h!]
	\centering
	\includegraphics[width=.48\textwidth]{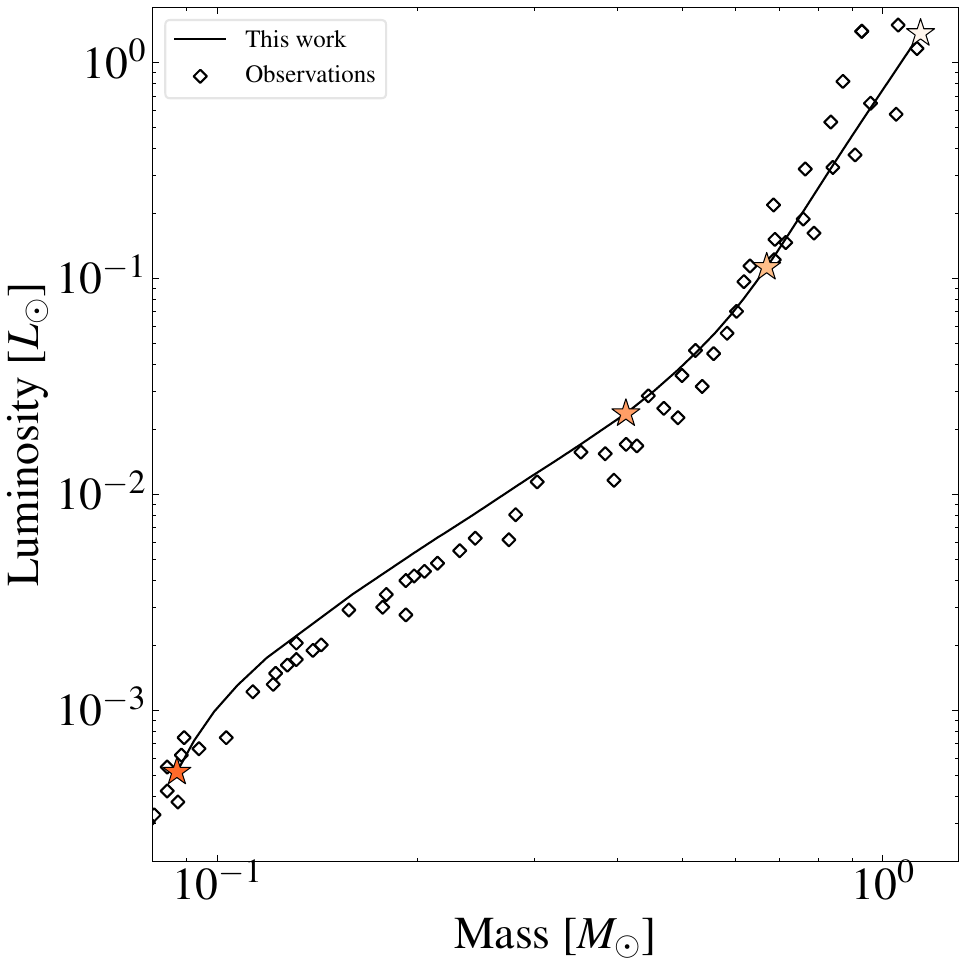}
    \includegraphics[width=.48\textwidth]{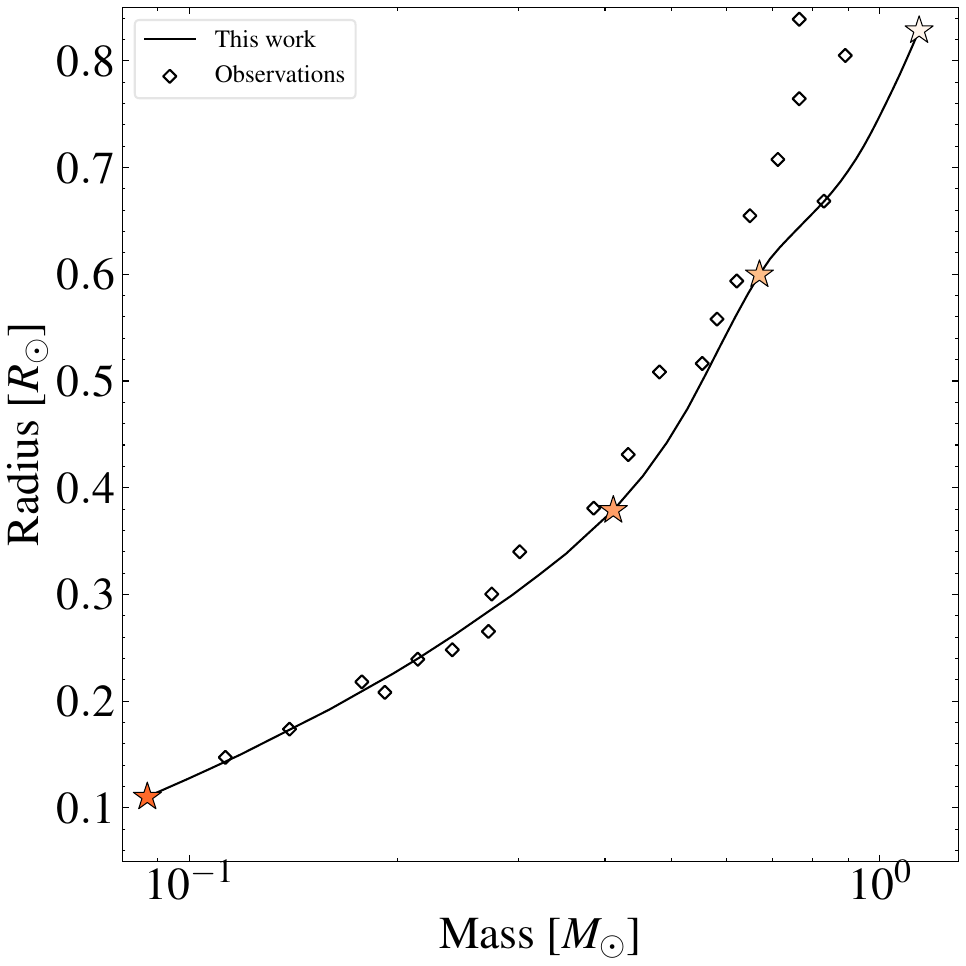}
    \includegraphics[width=.48\textwidth]{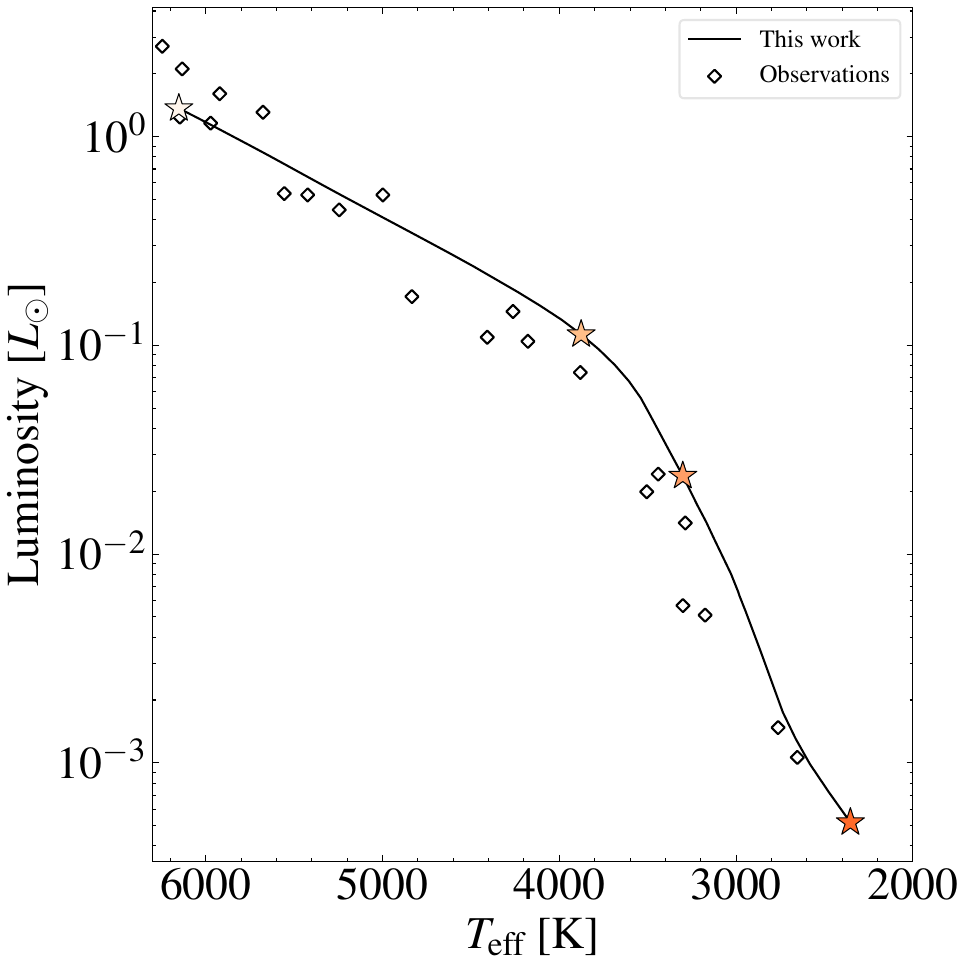}
	\caption{\textit{Upper left panel}: Mass-luminosity relation produced with our model. We compare with observations, taken from figure 3 of ref.~\cite{2018MNRAS.479.5491E} and figure 9 of~\cite{2019ApJ...871...63M}. \textit{Upper right panel}: Mass-radius relation produced with our model. The observations were taken from figure 2 of ref.~\cite{2018MNRAS.479.5491E}. \textit{Bottom panel}: Hertzsprung-Russell diagram produced with our model. The rhombus-shaped points correspond to observations of main sequence stars, taken from fig. 2.2.1 of ref.~\cite{2013sepp.book.....I}. In all panels, the colored star-shaped points correspond to the solutions with core temperature $T_c=1.5\times10^7,10^7,8\times10^6,4\times10^6$ K, with the color transitioning from a pale shade for the highest temperature to a deeper reddish-orange for the lowest, roughly reflecting the corresponding stellar surface colors.}
	\label{fig:StandardRelationsWhite}
\end{figure}

In fig.~\ref{fig:StandardRelationsWhite}, we present the mass-luminosity and mass-radius relations, together with the Hertzsprung-Russell (HR) diagram, predicted by our model. The corresponding observational data for main-sequence stars are shown for comparison. We find that the simplified stellar-structure treatment adopted here reproduces the observed relations at the level required for our purposes, despite not incorporating the full range of microphysical and evolutionary ingredients included in modern stellar-evolution codes. This agreement gives us confidence that the model provides a realistic description of the underlying stellar structure and is therefore suitable for investigating how the presence of a dark-sector component modifies the properties of stars.

In addition, we define four benchmark solutions that will be used throughout the rest of this article. These correspond to stars with core temperatures $T_c=1.5\times10^7,~10^7,~8\times10^6,~4\times10^6$ K, which are shown as colored star points in the figures. The solution with the lowest core temperature is depicted in reddish-orange, while the one with the highest core temperature appears pale white. This color coding approximately matches the observational expectation for each star based on its effective temperature. To derive color from effective temperature, we convert a Planck spectrum at $T_{\rm eff}$ into an sRGB color using the standard CIE 1931 color-matching functions~\cite{CIE1931}, which yields values in good agreement with those reported in Ref.~\cite{2021AN....342..578H}.

\begin{figure}[h!]
	\centering
	\includegraphics[width=.48\textwidth]{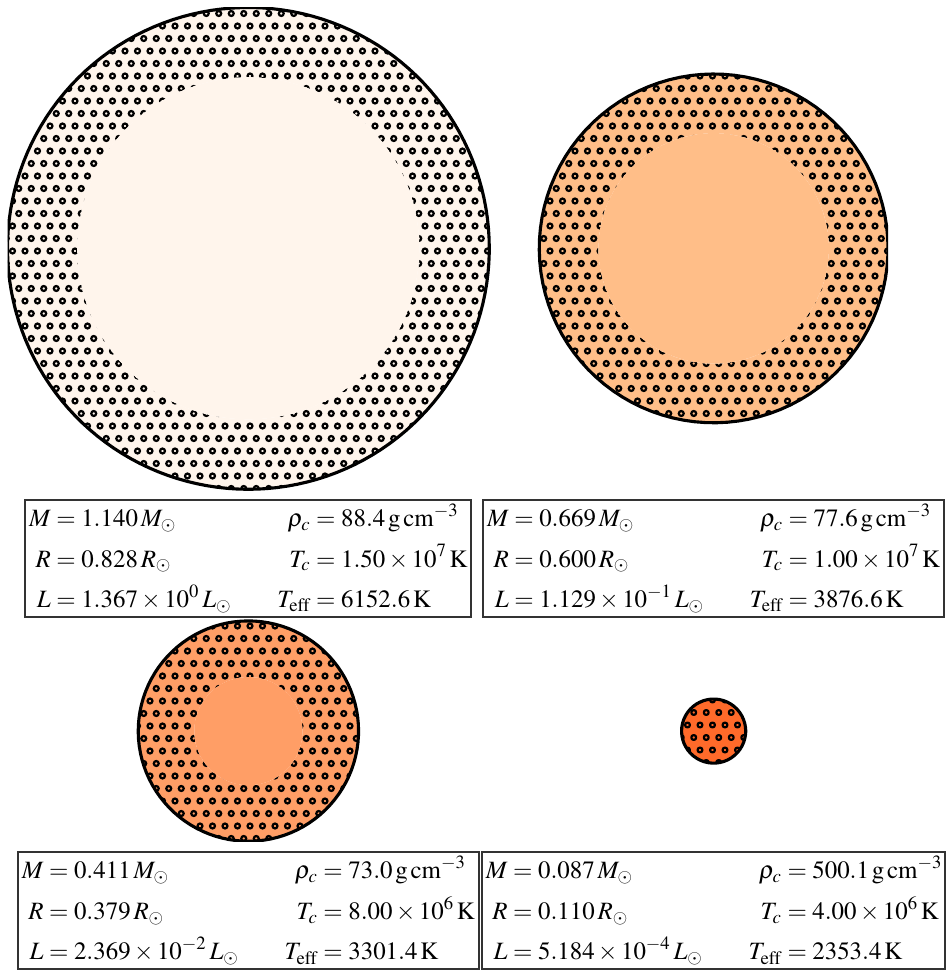}
    \includegraphics[width=.48\textwidth]{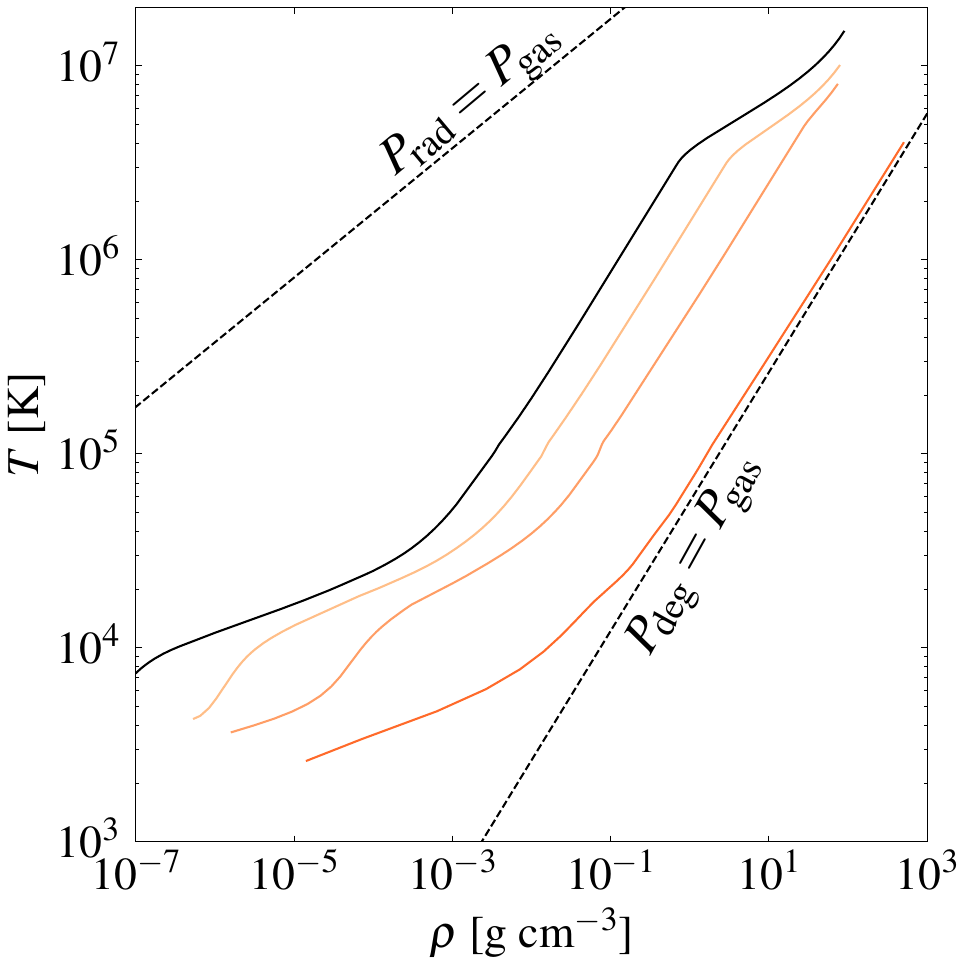}
	\caption{\textit{Left panel}: Two-dimensional sketches of the structure for the four colored points in fig.~\ref{fig:StandardRelationsWhite}. Each example shows the total mass, radius, luminosity, core density, and effective temperature. The dot-hatched region indicates where energy is transported by convection ($\nabla_{\rm rad} > \nabla_{\rm ad}$), while the plain region indicates radiative transport ($\nabla_{\rm ad} > \nabla_{\rm rad}$). The sketches follow the same color convention as in fig.~\ref{fig:StandardRelationsWhite} and are drawn to scale. \textit{Right panel}: Temperature–density profiles for the examples shown in the left panel. The black curve corresponds to $T_c=1.5\times10^7~\rm K$, while the other curves have the same color as their respective sketch. The two dashed lines delineate the regions where each component of the equation of state dominates: radiation dominated (left), ideal gas dominated (middle), and degeneracy dominated (right).}
	\label{fig:StandardSolutionsWhite}
\end{figure}

In the left panel of fig.~\ref{fig:StandardSolutionsWhite}, we present two-dimensional sketches (to scale) of the structure of our four benchmark stars, indicating the regions where energy transport is radiative (plain) and where it is convective (dot-hatched). Additional information on relevant quantities, such as mass and luminosity, is also provided. The right panel shows the temperature versus density profiles for the same stars, using the same color convention as in the left panel. For further comparison, we include the contours $P_{\rm rad}=P_{\rm gas}$ and $P_{\rm deg}=P_{\rm gas}$, based on the baryonic equation of state (EoS) in eq.~\eqref{eq:P_EoS}.

We note that all benchmark solutions lie well below the gas–radiation equality line, corresponding to a regime characteristic of low- to intermediate-mass main-sequence stars. In addition, our lowest-temperature benchmark point lies predominantly on the gas–degeneracy equality line, which corresponds to the lower mass limit of main-sequence stars. In this regime, electron degeneracy pressure begins to contribute meaningfully alongside gas pressure, supporting the star against gravitational collapse as nuclear burning becomes less efficient. Moving further below this equality line toward higher densities would lead to fully degenerate configurations such as brown dwarfs. Therefore, the fact that our lowest-temperature solution falls on this boundary, as well as the fact that this solution presents purely convective transport (as can be seen in the left panel) confirms that our model recovers the transition region between very-low-mass stars and brown dwarfs. Hence we are further justified in using our simplified model to study dark-sector influences near the stellar mass threshold.

\subsection{Dark Matter Admixed Star}
We now study how the stellar structure equations are modified by the presence of a DS. We assume that the two stars have the same center and, as mentioned earlier, the only coupling between the two is gravitational. The standard equations of stellar structure now become
\begin{equation}\label{eq:stellar_structure_DS}
\vcenter{\hbox{$
\begin{aligned}
\frac{dm_b}{dr} &= 4\pi r^2 \rho_b, \\
\frac{dP_b}{dr} &= -\frac{G(m_b+m_{\rm DS})\rho_b}{r^2}, \\
\frac{d\ell}{dr} &= 4\pi r^2 \rho_b \epsilon, \\
\frac{dT}{dr} &= \nabla \frac{T}{P_b}\frac{dP_b}{dr},
\end{aligned}
$}}
\qquad
\vcenter{\hbox{$
\begin{aligned}
\frac{dm_{\rm DS}}{dr} &= 4\pi r^2 \rho_{\rm DS}, \\
\frac{dP_{\rm DS}}{dr} &= -\frac{G(m_b+m_{\rm DS})\rho_{\rm DS}}{r^2}.
\end{aligned}
$}}
\end{equation}
Here, the subscript $b$ stands for baryonic matter and $\rm DS$ for the DS component. The radiative $\nabla$ is similarly modified by this addition into
\begin{equation}
    \nabla_{\rm rad}=\frac{3}{16\pi \alpha c}\frac{P_b\kappa}{T^4}\frac{\ell}{G(m_b+m_{\rm DS})}.
\end{equation}
We emphasize that in our model, the dark sector is included only as a source of gravity, not as a source of local pressure. Therefore, the hydrostatic balance for the baryonic gas is only influenced by the presence of DM in the gravitational acceleration term that contains the enclosed DS mass $m_{\rm DS}(r)$, while all pressure, opacity and radiative transport terms remain baryonic quantities (e.g. $P_b$, $\rho_b$, $\kappa(\rho_b,T)$, $\epsilon(\rho_b,T)$). This means that the DM doesn't contribute to either opacity or sources of energy.

\paragraph{Boundary conditions.} In this case, we have two additional  boundary conditions in the center for the DS component, and therefore the whole set of conditions at the center of both stars reads
\begin{align}
    &m_b(0)=\ell(0)=0,\, P_b(0)=P_{b,c}=P_b(\rho_{b,c},T_c,X,Z),\, T(0)=T_c,\cr
    &m_{\rm DS}(0)=0,\,P_{\rm DS}(0)=P_{\rm DS,c}=P_{\rm DS}(\rho_{\rm DS,c}),
\end{align}
where $\rho_{b,c}$ is the core density of baryonic matter and $\rho_{\rm DS,c}$ is the analogue for DM. The DS core pressure is then fixed by the DS equation of state. The small-radius expansion now becomes
\begin{align}
    m_b(r_\delta) &\approx \frac{4\pi}{3}r_\delta^3\rho_{b,c},\cr
    m_{\rm DS}(r_\delta) &\approx \frac{4\pi}{3}r_\delta^3\rho_{\rm DS,c},\cr
    P_b(r_\delta) &\approx P_{b,c}-\frac{2\pi G}{3}\rho_{b,c}(\rho_{b,c}+\rho_{\rm DS,c})r_\delta^2,\cr
    P_{\rm DS}(r_\delta) &\approx P_{\rm DS,c}-\frac{2\pi G}{3}\rho_{\rm DS,c}(\rho_{b,c}+\rho_{\rm DS,c})r_\delta^2,\cr
    \ell(r_\delta) &\approx \frac{4\pi}{3}r_\delta^3\rho_{b,c}\epsilon_c,\cr
    T(r_\delta) &\approx
    \begin{cases}
        T_c\,\left[1-\frac{\kappa_c\rho_{b,c}^2\epsilon_c}{2acT_c^4}r_\delta^2\right]^{\frac{1}{4}},\hspace{1.6cm}\text{if }\nabla_{\rm rad,c}\leq\nabla_{\rm ad,c}\cr
        T_c\,\mathrm{e}^{-\frac{2\pi G}{3}\frac{\rho_{b,c}(\rho_{b,c}+\rho_{\rm DS,c})}{P_{b,c}}\nabla_{\rm ad,c}r_\delta^2},\hspace{0.2cm}\text{if }\nabla_{\rm ad,c}\leq\nabla_{\rm rad,c}
    \end{cases}.
\end{align}
In direct analogy to the standard scenario, the radius of the baryonic component $R_b$ is determined by the photospheric boundary condition:
\begin{equation}
\label{eq:tau_boundary_DS}
\tau \approx \frac{\kappa P_b}{g} = \frac{2}{3}, \quad \text{where} \quad g = \frac{G (m_b(R_b)+m_{\rm DS}(R_b))}{R_b^{2}},
\end{equation}
which defines the total baryonic mass $M_b=m_b(R_b)$ and luminosity $L=\ell(R_b)$. For a fixed central temperature $T_c$, the central baryonic density $\rho_{b,c}$ is iterated until the additional condition $L = 4\pi R_b^{2} \sigma T(R_b)^{4}$ is satisfied, exactly as in the purely baryonic case. The dark matter component is integrated simultaneously, with its own central density $\rho_{\rm DS,c}$. Its radius $R_{\rm DS}$ is then implicitly defined by the condition $\rho_{\rm DS}(R_{\rm DS})=P_{\rm DS}(R_{\rm DS})=0$. Therefore, each choice of $\rho_{\rm DS,c}$ and $T_c$ yields a unique baryonic+DS model.

\paragraph{DM equation of state}
For simplicity we assume the DS to be a neutron star-like object in the sense that the temperature is much smaller than the chemical potential and effectively irrelevant for the DS EoS.
We consider two possibilities: a constant-density DS, defined by its core density $\rho_{\rm DS,c}$ and radius $R_{\rm DS}$, and a polytropic EoS $P_{\rm DS}=K\rho_{\rm DS}^{1+\frac{1}{n}}$. A polytropic EoS is anticipated in cases for example where DM is fermionic and degenerate. 

In the constant-density case, we explore several choices of density and radius. For the latter, we adopt $R_{\rm DS} = 0.05R_\odot, 0.1R_\odot, 0.15R_\odot,$ and $0.2R_\odot$, as well as the limiting case $R_{\rm DS} > R$, in which the DS fully encompasses the baryonic component. In this regime, the precise value of $R_{\rm DS}$ becomes irrelevant for the stellar structure equations~\eqref{eq:stellar_structure_DS}. We can see that the hydrostatic equilibrium equation for the baryons,
\begin{equation*}
    \frac{dP_b}{dr} = -\frac{G(m_b + m_{\rm DS})\rho_b}{r^2},
\end{equation*}
depends on the DS only through its contribution to the enclosed mass $m_{\rm DS}(r)$, which is determined by the constant density $\rho_{\rm DS}$. As long as the DS extends beyond the baryonic radius, its outer boundary does not affect the solution, and the only relevant parameter is the constant DM density $\rho_{\rm DS}$.

In the case of the polytrope, we take the representative cases $n=3/2$ and $n=1$. The first corresponds to degenerate, non-relativistic matter. This fixes the DM EoS to be of the form
\begin{equation}\label{eq:DM_EoS}
    P_{\rm DS}=\frac{1}{5}\left(\frac{3}{8\pi}\right)^{\frac23}\frac{h^2}{m_\chi^{8/3}}\rho_{\rm DS}^{5/3},
\end{equation}
where $m_\chi$ is the DM mass. The degenerate and non-relativistic conditions can only be satisfied for a given range of DM masses and densities. The quantum degeneracy condition $n_{\rm DS}\lambda_{\rm th}^3\gtrsim1$, where $\lambda_{\rm th}$ is the thermal de Broglie wavelength $\lambda_{\rm th}=h/\sqrt{2\pi m_\chi k_B T}$, imposes
\begin{equation*}
    \rho_{\rm DS}\gtrsim\frac{(2\pi k_B T)^{\frac32}m_\chi^\frac52}{h^3}=20.85~\frac{\mathrm{g}}{\mathrm{cm}^3}\left(\frac{m_\chi}{5~\mathrm {MeV}/c^2}\right)^\frac52\left(\frac{T}{10^7~\rm K}\right)^\frac32.
\end{equation*}
On the other hand, for degenerate fermions to be kinetically non-relativistic, we require $p_F\ll m_\chi c$, with $p_F=(h/2)(3 n_{\rm DS}/\pi)^\frac13$ the Fermi momentum. This yields the condition
\begin{equation*}
    \rho_{\rm DS}\ll\frac{8\pi m_\chi^4 c^3}{3h^3}=4.90\times10^6~\frac{\mathrm{g}}{\mathrm{cm}^3}\left(\frac{m_\chi}{5~\mathrm{MeV}/c^2}\right)^4.
\end{equation*}
Both conditions lead to the lower limit $m_\chi\gtrsim1.31~\mathrm{keV}/c^2\left(\frac{T}{10^7~\rm K}\right)$ for degenerate, non-relativistic DM.

On the other hand, the $n=1$ case is representative of the low density limit of self-interacting bosons and fermions. As relevant examples, we have the self-interacting complex scalar field model of Ref.~\cite{Colpi:1986ye} and the asymmetric fermionic DM model with self-interactions of Ref.~\cite{Kouvaris:2015rea}. Their EoSs  can be approximated in the low-density limit to be
\begin{equation}\label{eq:low_density_EoS}
\begin{cases}
    P\approx\frac{c^2\pi}{\rho_0}\rho^2,\hspace{2.41cm}\text{self-interacting bosons},\cr
    P\approx\frac{2\pi \hbar^3}{c}\frac{\alpha_D}{(m_{\chi}m_{\phi})^2}\rho^2,\hspace{0.9cm}\text{self-interacting fermions},
\end{cases}
\end{equation}
where $\rho_0=\left(2.92\times10^{21}\ \frac{\mathrm{kg}}{\mathrm{m}^3}\right)\lambda^{-1}\left(\frac{1\ \mathrm{GeV}/c^2}{m_\chi}\right)^{-4}$ with $\lambda$ the DM boson self-coupling and $m_\chi$ the dark boson mass. For fermions, $\alpha_D$ is the fine structure constant and $m_{\chi}$, $m_{\phi}$ are the dark fermion and dark mediator mass, respectively.

\begin{figure}[h!]
	\centering
	\includegraphics[width=.48\textwidth]{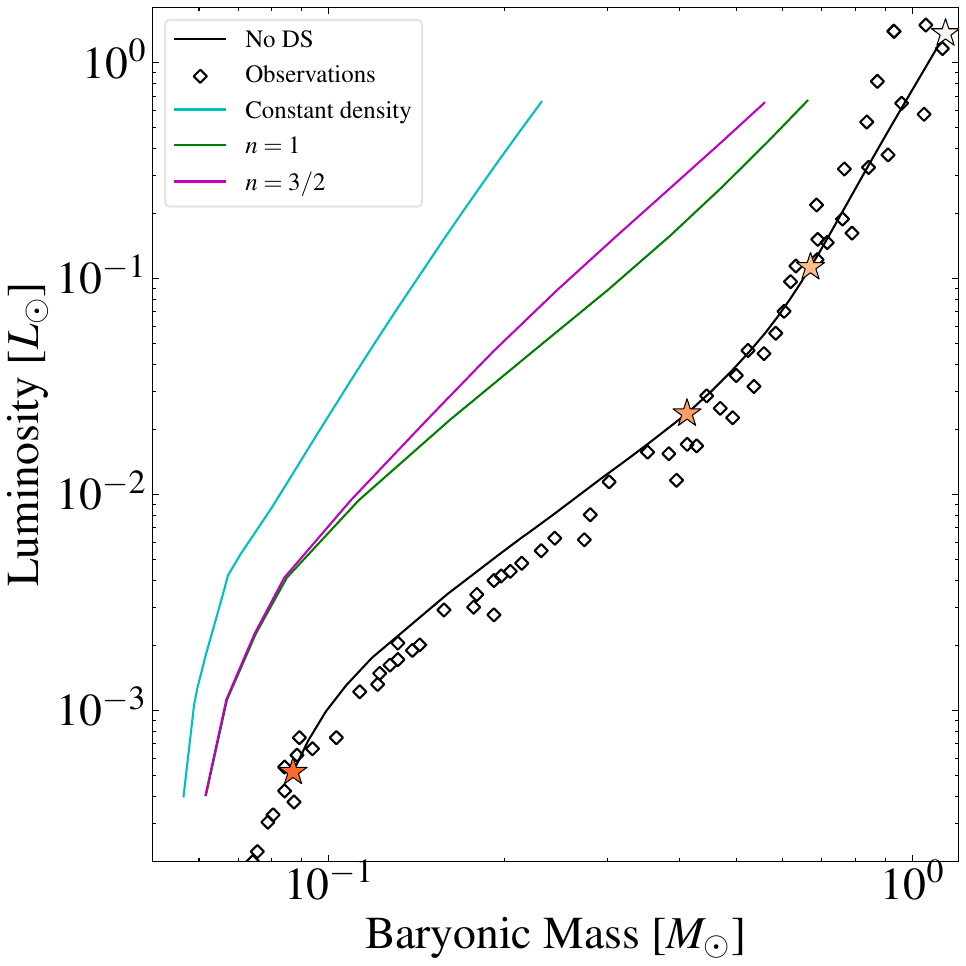}
    \includegraphics[width=.48\textwidth]{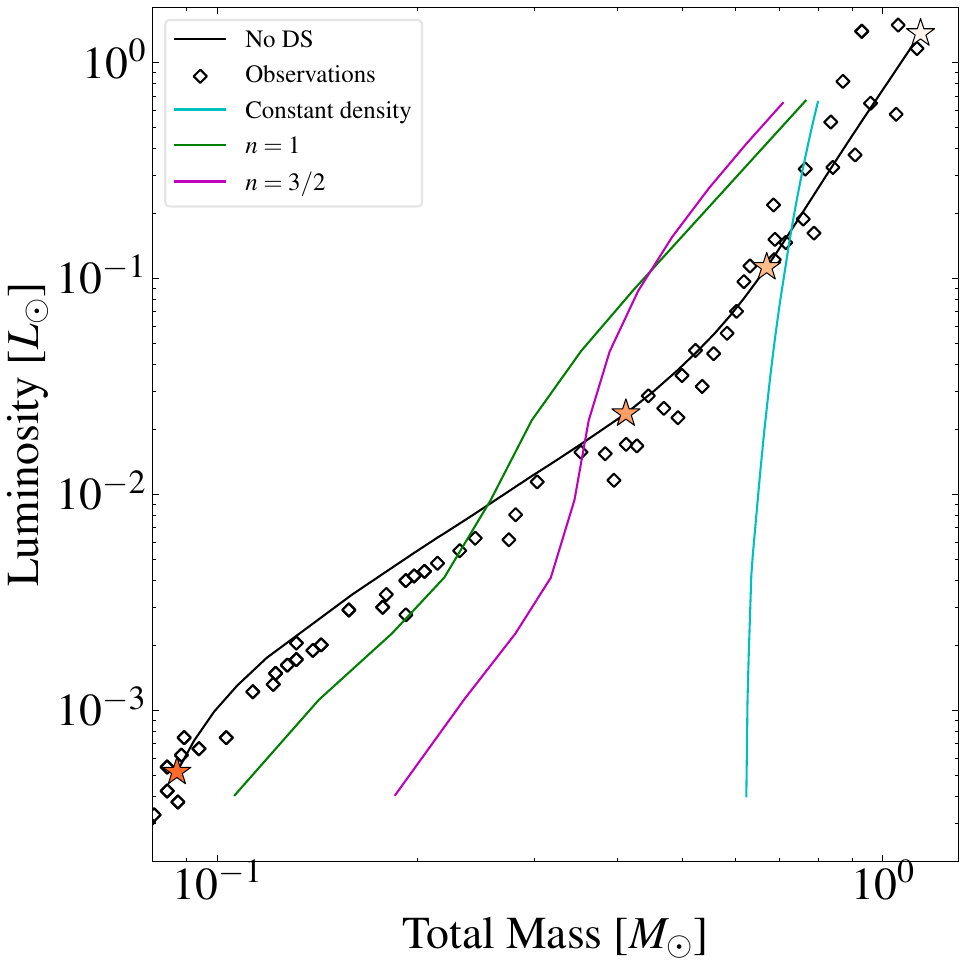}
    \includegraphics[width=.48\textwidth]{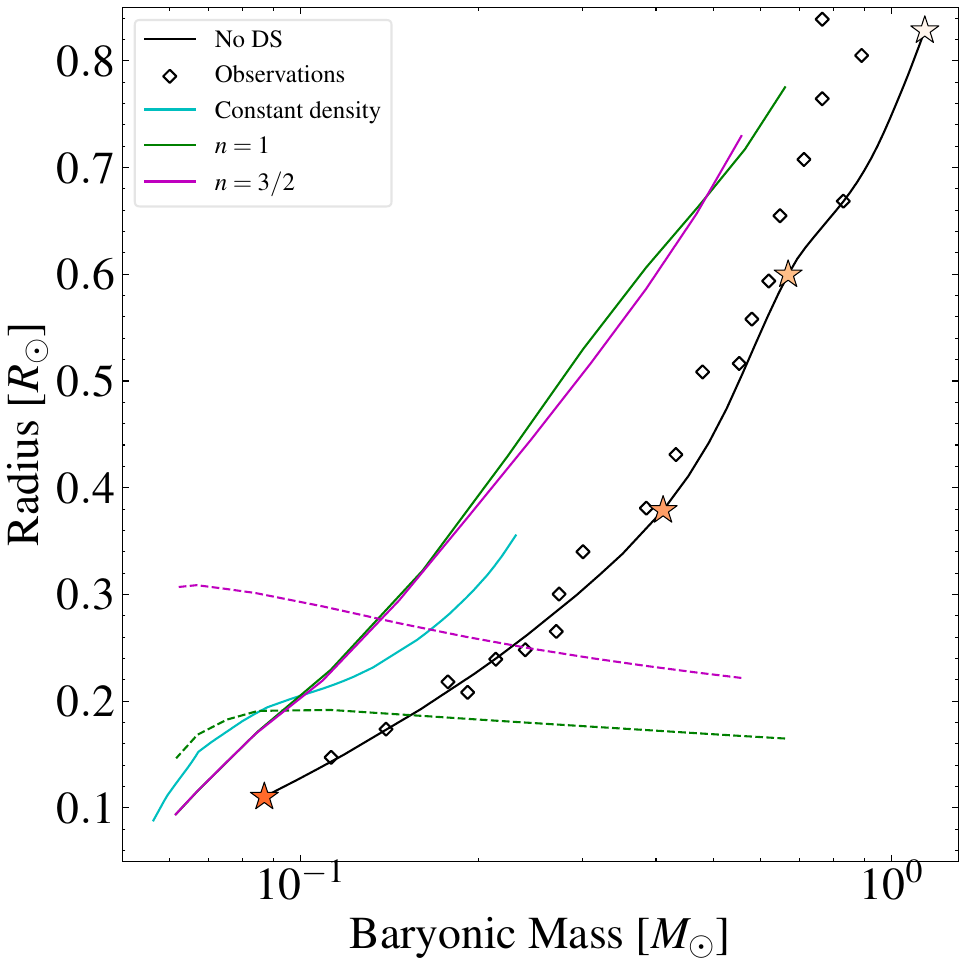}
    \includegraphics[width=.48\textwidth]{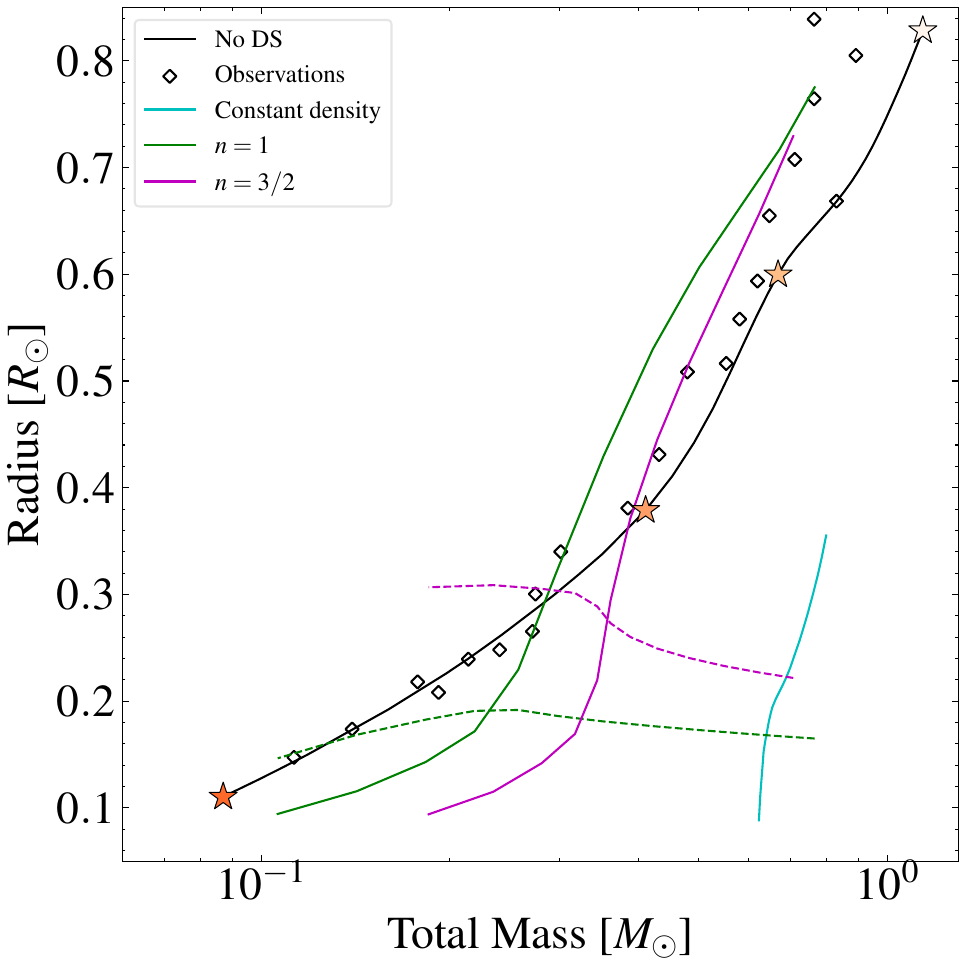}
	\caption{Different relations with the inclusion of DS. All DSs have a fixed core density of $\rho_{\rm DS,c}=100~\mathrm{g}~\mathrm{cm}^{-3}$ and polytropic constant $K=3\times10^{16}~\mathrm{Pa}/(150~\mathrm{g}~\mathrm{cm}^{-3})^{(n+1)/n}$. For the case of a DS with constant density, we choose a radius $R_{\rm DS}=0.2R_\odot$. The colored regions and star-shaped points follow the same conventions as in fig.~\ref{fig:StandardRelationsWhite}. \textit{Upper panels}: Mass-luminosity relation as a function of baryonic mass (left) and total mass (right). \textit{Bottom panels}: Mass-radius relation as a function of baryonic mass (left) and total mass (right). The solid curves are the baryonic radius, while the dashed curves represent the radius of the DS.}
	\label{fig:DSRelations}
\end{figure}

In fig.~\ref{fig:DSRelations} we present mass‑luminosity and mass‑radius relations for the DM EoSs under consideration. We consider a polytropic constant $K=3\times10^{16}~\mathrm{Pa}/(150~\mathrm{g}~\mathrm{cm}^{-3})^{(n+1)/n}$, which is motivated by the core conditions of a solar-like star with a core density of $\rho\sim150~\mathrm{g}~\mathrm{cm}^{-3}$ and core pressure $P\sim3\times10^{16}~\mathrm{Pa}$. This choice fixes the overall scale of the DS EoS to a value that is neither negligibly small nor overwhelmingly large compared to the baryonic core. In terms of microphysical parameters, it would correspond to $\lambda^{-1/4}(m_\chi c^2/1~\mathrm{GeV})=0.00292$ for self-interacting bosons and to $(\alpha_D/0.1)(m_{\chi}c^2/1~\mathrm{GeV})^{-1}(m_{\phi}c^2/1~\mathrm{eV})^{-1}=7.4\times10^{-5}$ for self-interacting fermions according to eq.~\eqref{eq:low_density_EoS}.

We have also examined the polytropic DM with $n=3/2$ using the polytropic constant given by eq.~\eqref{eq:DM_EoS} for DM masses $m_\chi=2~\mathrm{MeV}/c^2$ and $m_\chi=5~\mathrm{MeV}/c^2$. However, the resulting baryonic mass‑luminosity and baryonic mass‑radius relations show no difference from the constant‑density case, so we do not show them separately in the figure.

The inclusion of a dark component alters the stellar structure in a way that is clearly reflected in the mass-luminosity relation shown in the upper left panel of the figure. The presence of DM increases the luminosity of the baryonic star. DM increases the gravitational pull leading typically to larger central densities and temperatures in order to maintain hydrostatic equilibrium. In turn it increases the amount of energy produced by fusion and thus the luminosity output. As we will demonstrate, an immediate consequence of this is that the minimum baryonic mass required to support a main-sequence star can be pushed below the  usual hydrogen-burning limit of $\sim 0.085\,M_\odot$. Despite the smaller baryonic mass, the presence of DM increases the central density and temperature and therefore sustains the hydrogen fusion.

\begin{figure}[h!]
	\centering
	\includegraphics[width=.48\textwidth]{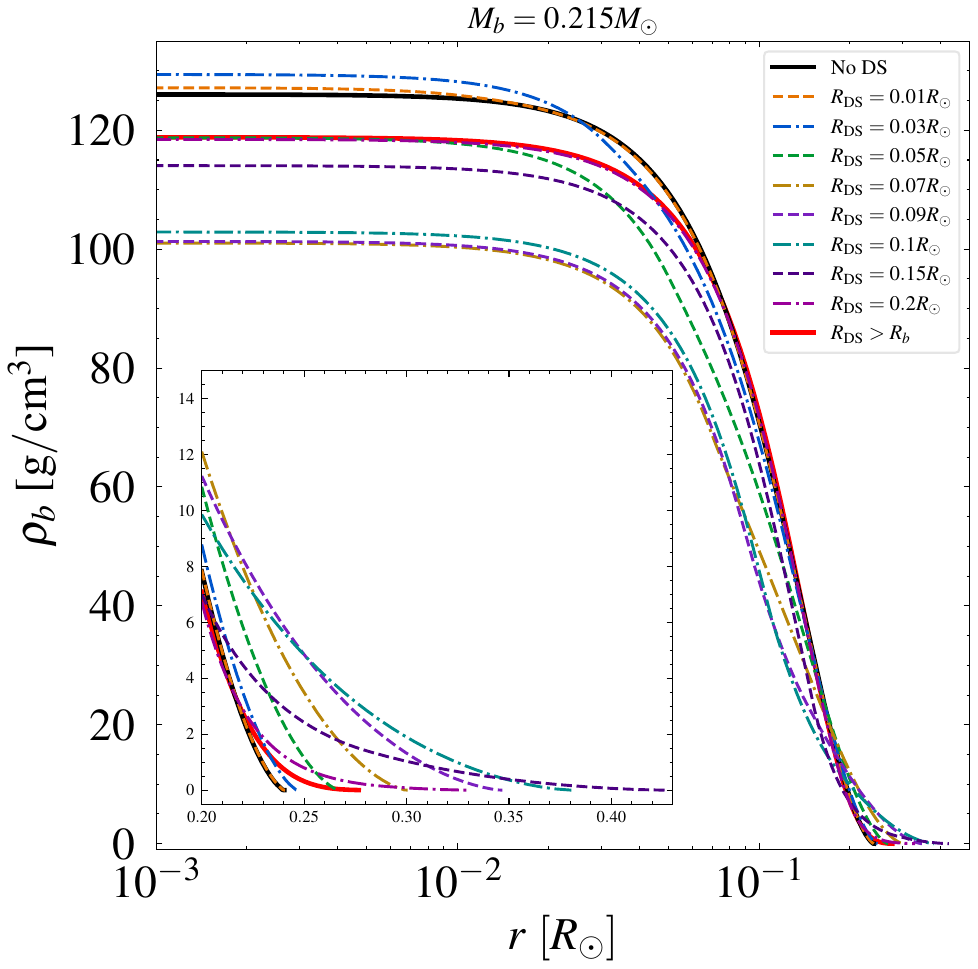}
    \includegraphics[width=.48\textwidth]{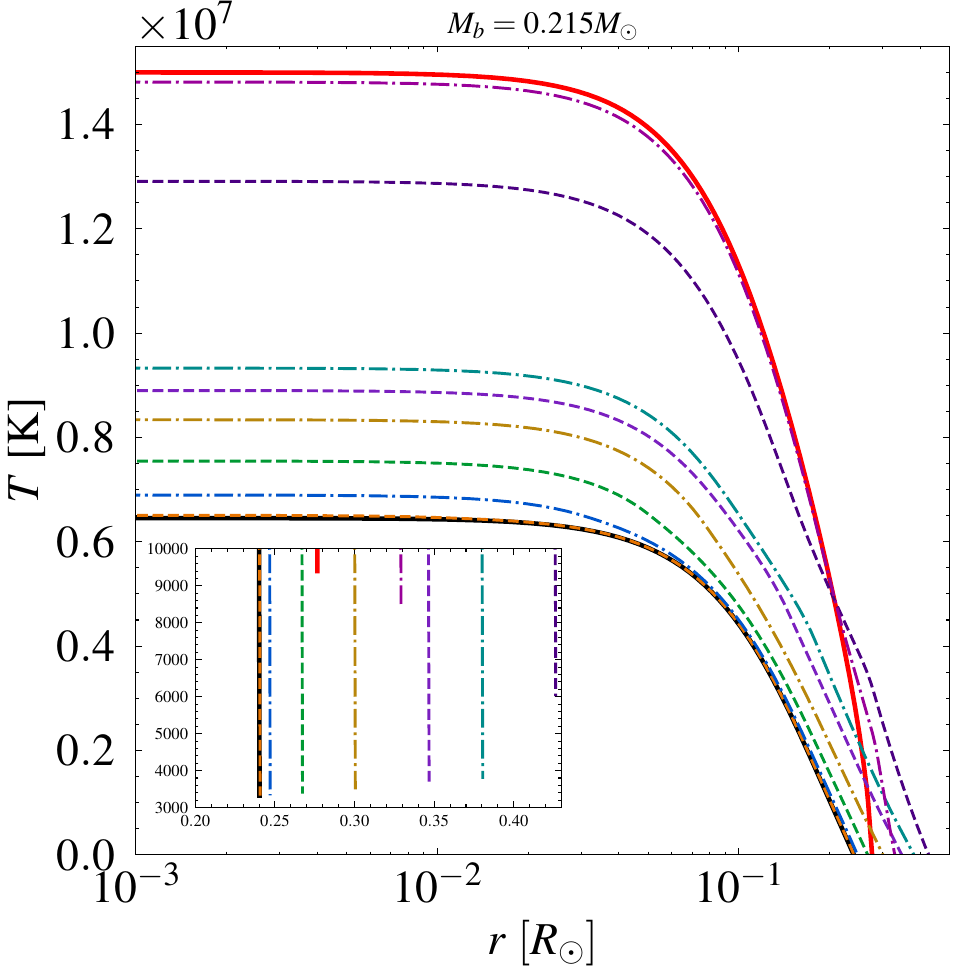}
    \includegraphics[width=.48\textwidth]{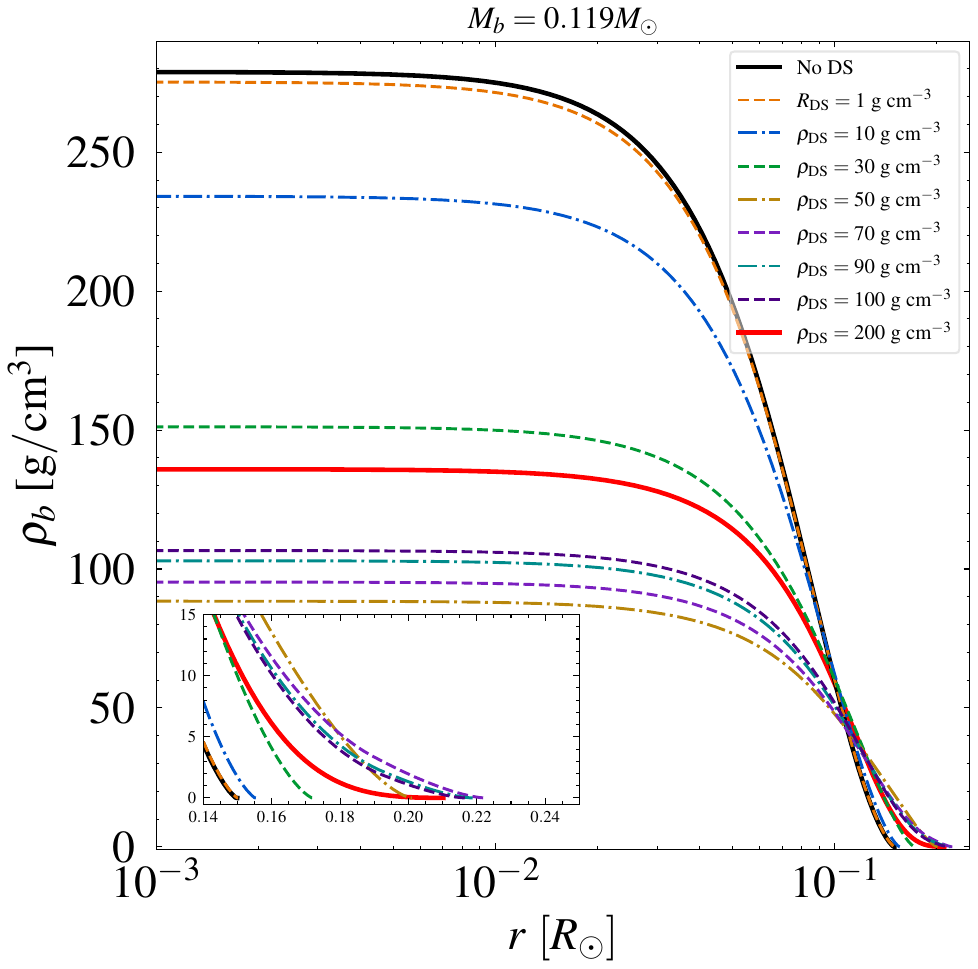}
    \includegraphics[width=.48\textwidth]{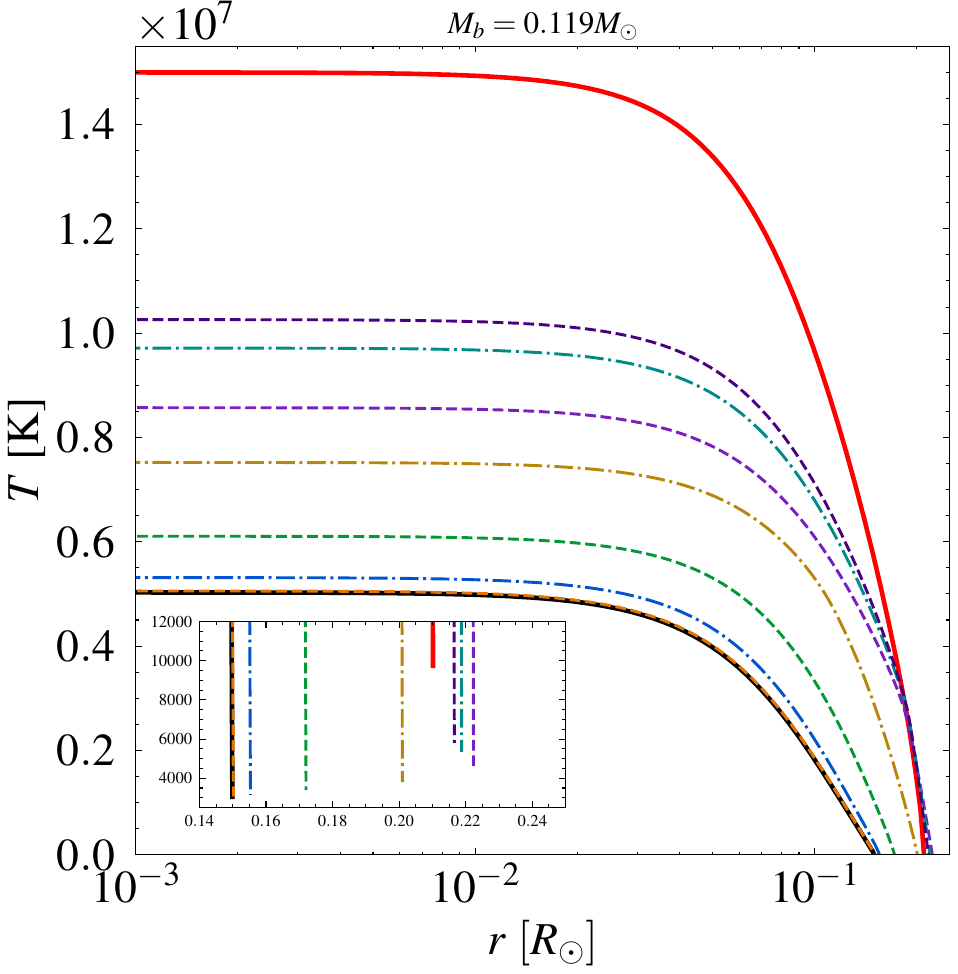}
	\caption{\textit{Left panels:} Baryonic density profiles with a fixed constant DS density (colored curves) $\rho_{\rm DS}=100~\mathrm{g}~\mathrm{cm}^{-3}$ for various DS radii (\textit{Upper panel}) and fixed DS radius (that exceeds the baryonic radius) for various constant DS densities (\textit{Bottom panel}). \textit{Right panels:} Temperature profiles of the solutions of the left panels.
    The baryonic mass is kept fixed ($0.215M_\odot$ in the upper panels and $0.119M_\odot$ in the bottom panels). In all panels, the DM-free solution is shown as a solid black line for comparison.}
	\label{fig:DensityProfiles}
\end{figure}

 The bottom left panel of fig.~\ref{fig:DSRelations} shows the baryonic radius (solid curves) and the DS radius (dashed curves) as functions of baryonic mass. For a given baryonic mass, the radius of the admixed star is larger than that of a pure baryonic star (with no DM) of the same mass, indicating  that the baryonic matter expands when admixed with DM. This result may appear contradictory at first, since adding extra mass (the DM) would increase the total gravitational force, which one might naively expect to compress the star. However, the baryons must remain in hydrostatic equilibrium. The added DS deepens the gravitational potential, forcing the baryonic gas to become hotter in order to provide the necessary pressure support. This increases the luminosity as we mentioned and as a consequence, the envelope of the star expands. The situation is somewhat analogous to the case of red giants where an inert helium core does not produce substantial energy (as DM in our case). However, it increases the temperature of the hydrogen-burning shell and therefore the luminosity. The additional energy released shifts the equilibrium configuration toward a larger radius (as it occurs in our case). The analogy can only be taken so far. In red giants the helium core participates in the pressure support in the form of Fermi pressure in contrast to the DM case where its presence affects only the gravitational potential and provides no pressure for the baryonic component.  
Therefore  this expansion in the admixed star is not monotonic: as more DM is added, the competition between the deepening potential and the rising thermal support can eventually reverse, causing the baryonic radius to reach a maximum and then decrease. See fig.~\ref{fig:DensityProfiles} for concrete examples and a detailed discussion of this balance.

Finally, we point out that for the case following the polytropic constant of eq.~\eqref{eq:DM_EoS} $m_\chi=2~\mathrm{MeV}/c^2$ and a fixed core density $\rho_{\rm DS,c}=100~\mathrm{g}~\mathrm{cm}^{-3}$, we obtain a DS mass $M_{\rm DS}\sim1.7\times10^{4}M_\odot$ and radius $R_{\rm DS}\sim11.3R_\odot$. For the other case $m_\chi=5~\mathrm{MeV}/c^2$ with the same core density, we find $M_{\rm DS}\sim430M_\odot$ and radius of $R_{\rm DS}\sim3.3R_\odot$. These values are substantial, and one might worry that such massive dark objects are ruled out by existing astrophysical constraints. We now discuss why this is not the case.

Although a DS with $M_{\rm DS}\gtrsim 10^3\,M_\odot$ may appear to be in tension with existing astrophysical bounds on DM in the form of compact objects, the strongest limits in this mass range are not absolute and depend on several uncertain astrophysical assumptions. The main constraints usually discussed in the literature come from dynamical heating of compact stellar systems such as Eridanus~II, X-ray emission from accretion and binary formation, and CMB limits on energy injection by accreting compact objects~\cite{Cirelli:2024ssz}. The Eridanus II argument, for example, states that repeated gravitational encounters between massive compact halo objects (MACHOs) and stars in a compact cluster would heat and dissolve the cluster over gigayear timescales. Since the cluster in Eridanus II is still present, MACHOs cannot constitute most of the DM~\cite{Brandt:2016aco}. The Eridanus~II constraint is nevertheless subject to several astrophysical uncertainties. Firstly, the inferred heating rate depends sensitively on the assumed age of the cluster, that is, if the cluster formed more recently than in the fiducial estimate, then there may have been insufficient time for significant expansion. Secondly, the argument depends on the present-day structure of the host halo, in particular on the local DM density and velocity dispersion. A cored halo, or a larger velocity dispersion, would reduce the rate of gravitational encounters and weaken the constraint. Thirdly, the cluster may not have remained continuously embedded in the densest region of the halo, so the cumulative heating could be substantially smaller than in the idealized estimate. Finally, the usual bounds are derived for a monochromatic population of compact objects, whereas the present dark-sector configurations are extended and can differ substantially from the point-mass approximation made in the Eridanus II analysis.

X-ray and accretion constraints are also less directly applicable to the present scenario. For a population of $\sim1000\,M_\odot$ black holes moving through the interstellar medium, Bondi-like accretion can produce observable X-ray emission, while accreting binaries may appear as ultraluminous X-ray sources. The absence of a large population of such objects is therefore often used to constrain compact objects in this mass range. However, the DSs considered here differ fundamentally from black holes. They are extended self-gravitating configurations of dark-sector particles and possess neither an event horizon nor a baryonic surface. Their compactness is not necessarily large. Consequently, the standard accretion models employed in deriving X-ray limits for black holes cannot be directly transferred to our case. In particular, the efficiency with which baryonic material can be captured by, thermalized within, and radiatively cool inside an extended dark object is highly model dependent. Although baryonic gas is gravitationally attracted toward the dark star, the resulting luminosity could be substantially lower than in the black-hole case because the mechanisms responsible for converting gravitational binding energy into observable radiation are expected to differ. As an example, we have previously performed a quantitative assessment of the associated X-ray signal in ref.~\cite{Kamenetskaia:2022lbf} for a scenario with a bosonic DS. We found that the X-ray luminosity could be sizable only for a very short amount of time during an outburst. However, both the luminosity and timescales are much shorter than the values used for black-hole X-ray constraints~\cite{Inoue:2017csr,Manshanden:2018tze}. For this reason, existing X-ray constraints derived for accreting black holes should not be regarded as directly excluding the DS models considered here.

Finally, CMB distortion limits arise from energy injection by accreting primordial black holes before recombination, which alters the ionization history and leaves imprints on Planck measurements~\cite{Ricotti:2007au,Poulin:2017bwe}. Once again, they assume compact, accreting black holes, which differ substantially from DSs, as mentioned previously for the X-ray case.

\begin{figure}[h!]
	\centering
	\includegraphics[width=.48\textwidth]{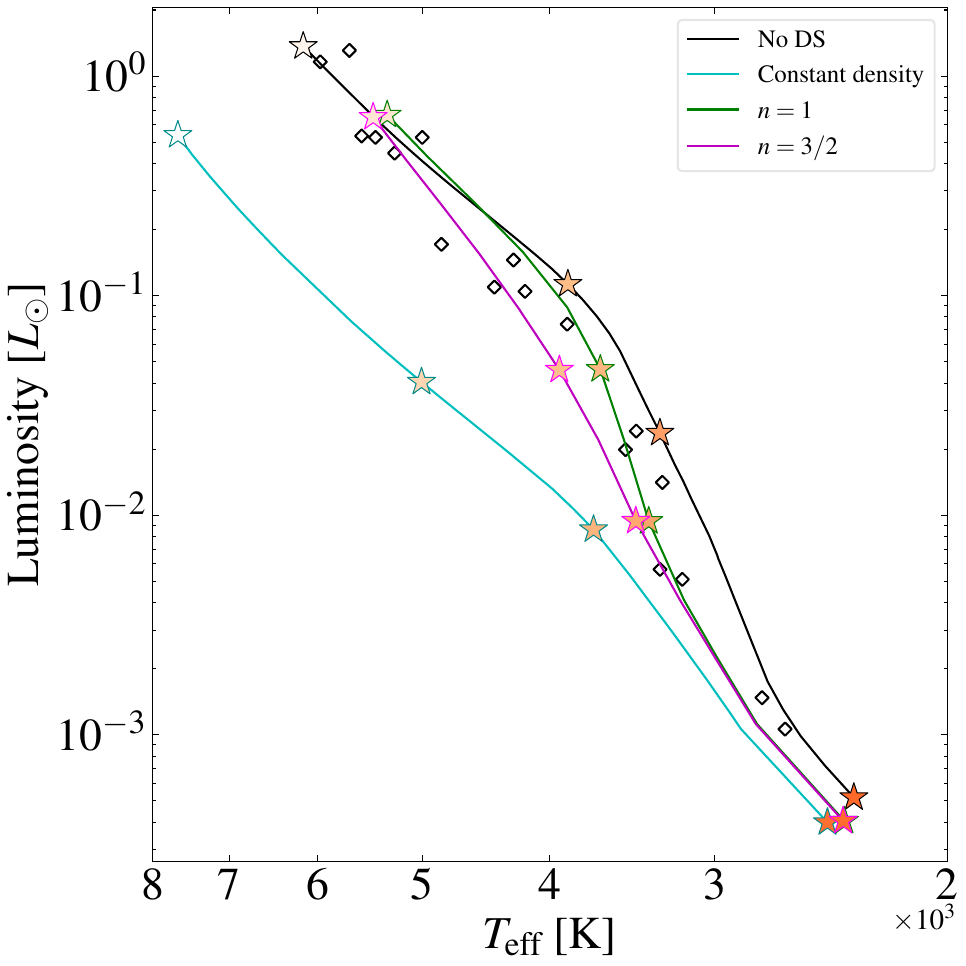}
	\caption{HR diagram for different DS models. All DSs have a fixed core density of $\rho_{\rm DS,c}=100~\mathrm{g}~\mathrm{cm}^{-3}$ and polytropic constant $K=3\times10^{16}~\mathrm{Pa}/(150~\mathrm{g}~\mathrm{cm}^{-3})^{(n+1)/n}$. The constant density DS has a radius $R_{\rm DS}=0.2R_\odot$. The rhombus-shaped points are the same as in the bottom panel of fig.~\ref{fig:StandardRelationsWhite}.}
	\label{fig:HRDiagramDS}
\end{figure}

In fig.~\ref{fig:HRDiagramDS}, we show the HR diagram for the same cases used in fig.~\ref{fig:DSRelations}. This figure shows that the addition of DM modifies the surface properties of the baryonic component in a non-trivial way. Since the effective temperature is related to luminosity and radius through the Stefan-Boltzmann law $L=4\pi R^2 \sigma T_{\rm eff}^4$, the change in $T_{\rm eff}$ reflects a competition between the DM-induced increase in luminosity and the simultaneous expansion of the baryonic radius. Typically, DM tends to heat the baryonic interior and enhance the luminosity, but it also increases the stellar radius. Namely, if the luminosity increase dominates, the effective temperature  becomes larger, while if the radius increase is dominant, the effective temperature drops. In the low-mass regime (low-luminosity, as per fig.~\ref{fig:DSRelations}), where degeneracy pressure becomes more important and the baryonic EOS is less sensitive to temperature, the response is weaker and the shift in $T_{\rm eff}$ is smaller. At larger baryonic masses (large luminosity), where the EoS is more gas-pressure dominated, the DM-induced thermal expansion becomes more pronounced and the radius increase can outweigh the luminosity enhancement, leading in some cases to  lower $T_{\rm eff}$ than in the no-DS case.

\begin{figure}[h!]
	\centering
	\includegraphics[width=.48\textwidth]{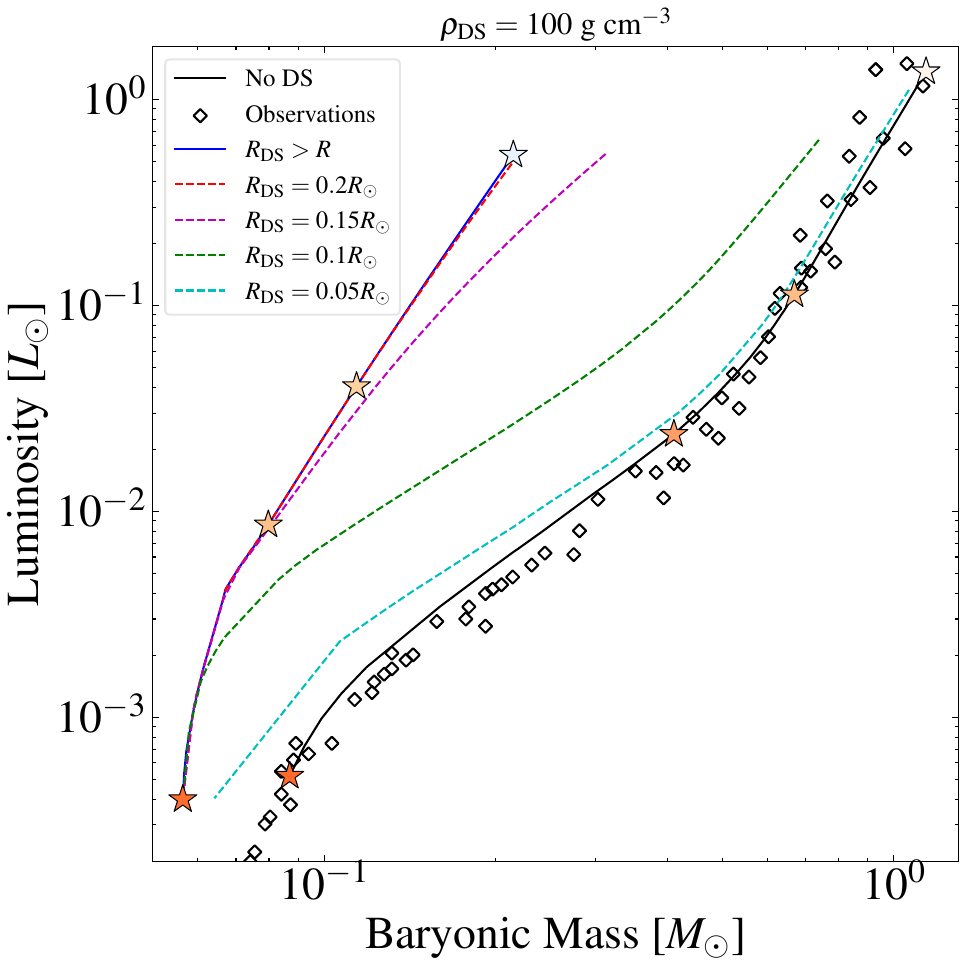}
    \includegraphics[width=.48\textwidth]{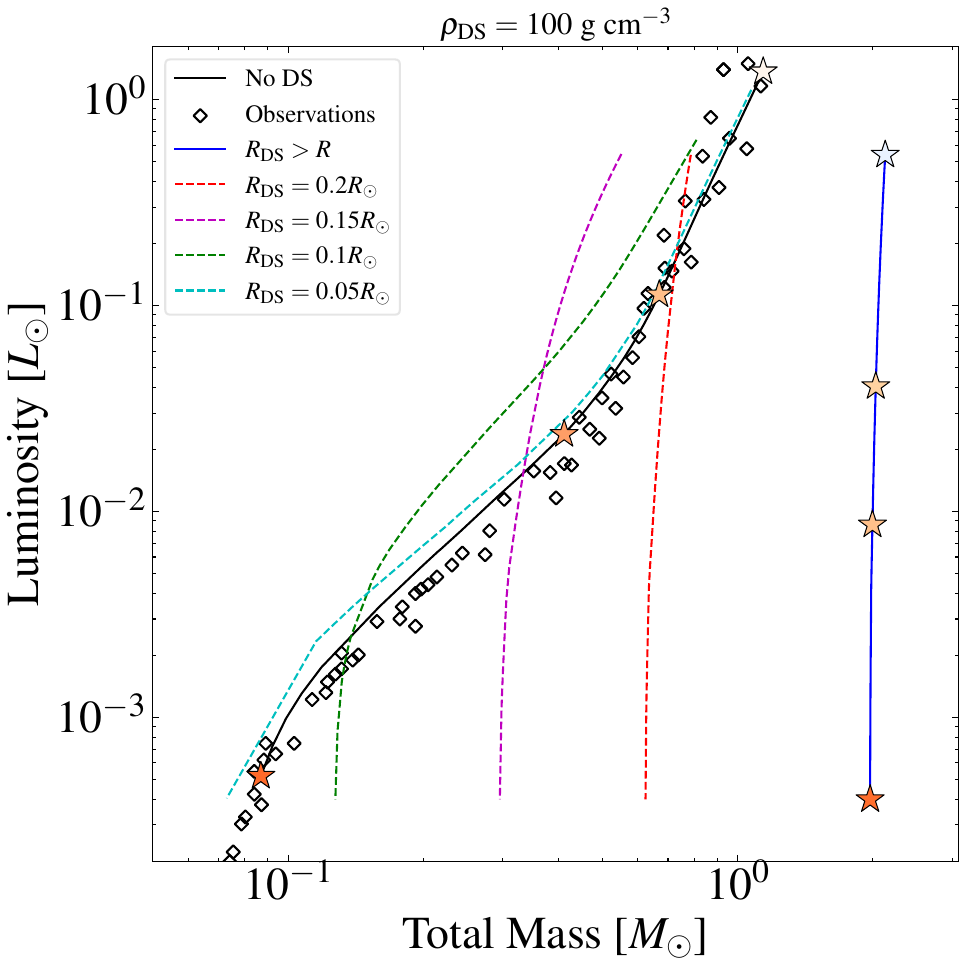}
    \includegraphics[width=.48\textwidth]{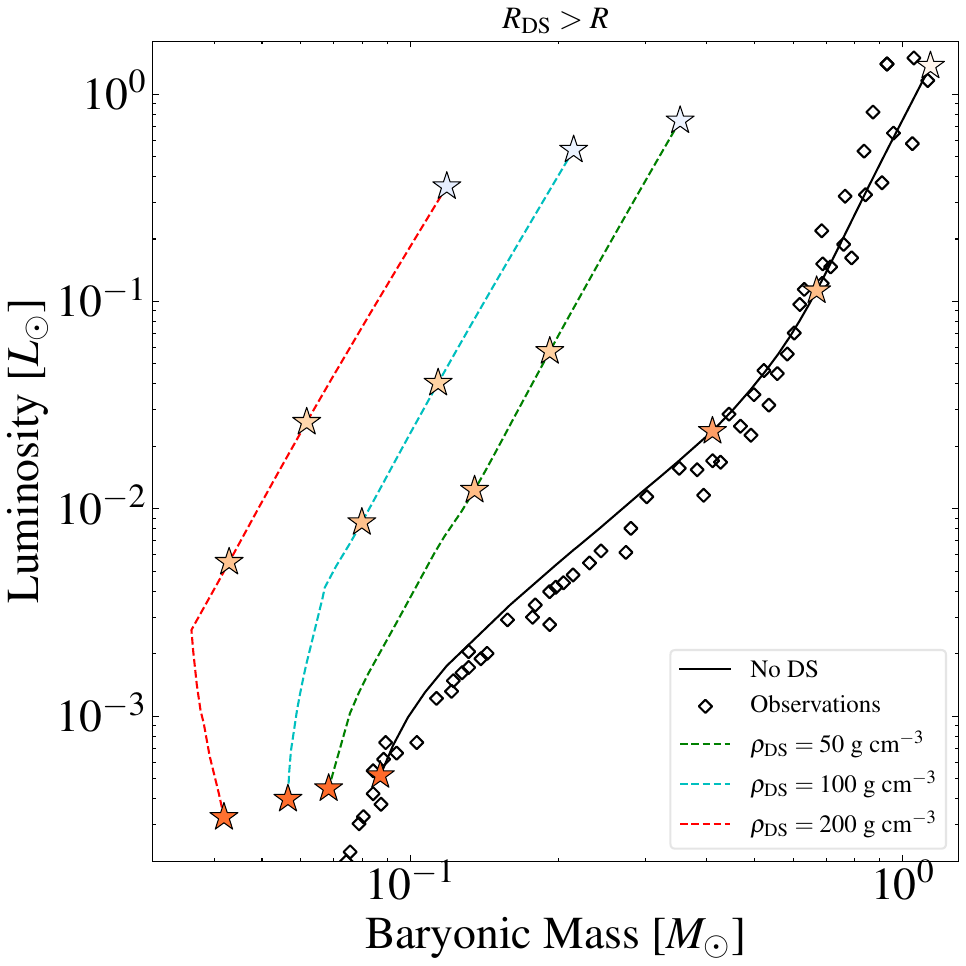}
	\caption{\textit{Upper panels}: Mass-luminosity relation for a constant density DS ($\rho_{\rm DS}=100~\mathrm{g}~\mathrm{cm}^{-3}$) with varying radii $R_{\rm DS}$. The left panel shows only the baryonic mass, while the right panel presents the total mass. \textit{Bottom panel}: Baryonic mass-luminosity relation for different choices of density of the constant-density DS. The DS radius is chosen such that it always encompasses the totality of the baryonic component.}
	\label{fig:Comparisons}
\end{figure}

In fig.~\ref{fig:Comparisons}, we show different mass-luminosity relations for constant density DSs. In the upper panels, we keep the core density fixed to $\rho_{\rm DS}=100~\mathrm{g}~\mathrm{cm}^{-3}$ and vary the radius $R_{\rm DS}$ to estimate the impact on the baryonic star's properties as a function of the size of the DS. The solid blue curve corresponds to the limiting case where the DS is always larger than the baryonic part. We note, in the same way as we did for the solutions in fig.~\ref{fig:DSRelations}, that the addition of DM leads to larger luminosity for fixed baryonic mass. As expected, the magnitude of this effect increases with the size and mass of the DS, leading to progressively larger deviations from the standard mass-luminosity relation. 

Another notable feature is the convergence of the dashed curves toward the limiting solid blue curve at low baryonic masses. This trend becomes particularly evident for configurations with $R_{\rm DS}\gtrsim0.1R_\odot$, where the two curves eventually coincide. The reason for this behavior is that, as the baryonic mass decreases, the baryonic radius also decreases and eventually becomes smaller than the DS radius. Once the baryonic component is fully embedded within the DS, the baryons experience essentially the same DM environment regardless of their exact radius. Consequently, the stellar structure approaches the limiting configuration represented by the solid blue curve, causing the mass-luminosity relations to merge.

Finally, in the lower panel we present the baryonic mass-luminosity relation for a constant-density DS with three different densities. Here we can once again observe that the increase in the luminosity as a function of the baryonic mass directly depends on the DS density, with the highest density solutions leading to the largest luminosity. Furthermore, the minimum baryonic mass limit also explicitly depends on this quantity. As one can see, the minimum baryonic mass for hydrogen fusion drops from the standard limit of $M_{\rm min}\sim0.085M_\odot$ (no DM present)  to $M_{\rm min}\sim0.045M_\odot$ for $\rho_{\rm DS}=200~\mathrm{g}~\mathrm{cm}^{-3}$.

\begin{figure}[h!]
	\centering
	\includegraphics[width=.98\textwidth]{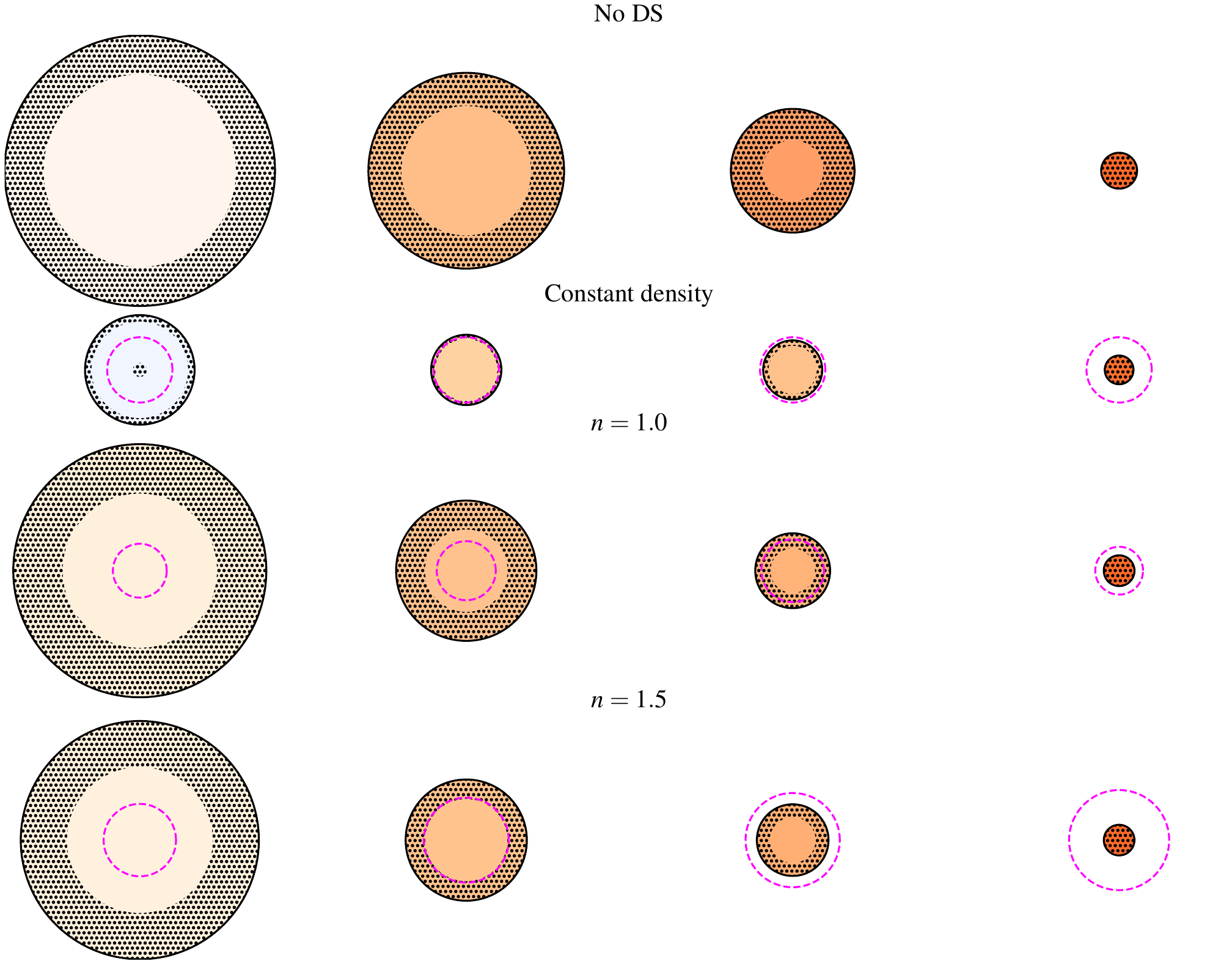}
	\caption{Two-dimensional sketches of the structure for solutions with core temperature $T_c=1.5\times10^7,10^7,8\times10^6,4\times10^6$ K. The upper panels show solutions with no DS, the second row assumes a constant-density DS with radius $R_{\rm DS}=0.2R_\odot$, while the two bottom panels consider a DS with $n=1$ and $n=3/2$, respectively with polytropic constant $K=3\times10^{16}~\mathrm{Pa}/(150~\mathrm{g}~\mathrm{cm}^{-3})^{(n+1)/n}$. All DSs are assumed to have a core density of $\rho_{\rm DS,c}=100~\mathrm{g}~\mathrm{cm}^{-3}$. The color and hatch conventions follow the same rules of the left panel of fig.~\ref{fig:StandardSolutionsWhite}. Additionally, the dashed magenta circles in the bottom panels represent the size of the DS. Information on each solution is provided in table~\ref{table:2d_comparison}.}
	\label{fig:2DSolutions_comparison}
\end{figure}

\begin{table}[h!]
\caption{Total baryonic mass $M_b$, total DS mass $M_{\rm DS}$, baryonic radius $R_b$, DS radius $R_{\rm DS}$, core baryonic density $\rho_{b,c}$, total luminosity $L$ and effective temperature $T_{\rm eff}$ for the solutions of fig.~\ref{fig:2DSolutions_comparison}. All DSs have a core density of $\rho_{\rm DS,c}=100~\mathrm{g}~\mathrm{cm}^{-3}$.}
\label{table:2d_comparison}
\setlength{\tabcolsep}{2pt}
\begin{tabular}{cccccccc}
\hline \hline
                       & $M_b$ $[M_\odot]$ & $M_{\rm DS}$ $[M_\odot]$ & $R_b$ $[R_\odot]$ & $R_{\rm DS}$ $[R_\odot]$ & $\rho_{b,c}$ $[\mathrm{g}~\mathrm{cm}^{-3}]$ & $L$ $[L_\odot]$ & $T_{\rm eff}$ [K] \\
\hline
No DS & & & & & & & \\
\hline
$T_c=15$ MK & $1.140$ & $-$ & $0.828$ & $-$ & $88.4$ &  $1.37$ &  $6152.6$ \\
$T_c=10$ MK   & $0.669$ & $-$ & $0.600$ & $-$ & $77.6$ &  $1.13\times10^{-1}$ &  $3876.6$ \\
$T_c=8$ MK    & $0.411$ & $-$ & $0.379$ & $-$ & $73.0$ &  $2.37\times10^{-2}$ &  $3301.4$ \\
$T_c=4$ MK    & $0.087$ & $-$ & $0.110$ & $-$ & $500.1$ &  $5.18\times10^{-4}$ &  $2353.4$ \\
\hline
Constant density                &                   &                          &                   &                          &                                              &                 &                       \\
\hline
$T_c=15$ MK & $0.219$ & $0.567$ & $0.336$ & $0.2$ & $118.8$ &  $0.54$ &  $7657.6$ \\
$T_c=10$ MK   & $0.114$ & $0.567$ & $0.215$ & $0.2$ & $105.8$ &  $4.04\times10^{-2}$ &  $5005.0$ \\
$T_c=8$ MK    & $0.080$ & $0.567$ & $0.181$ & $0.2$ & $99.4$ &  $8.61\times10^{-3}$ &  $3706.8$ \\
$T_c=4$ MK    & $0.056$ & $0.567$ & $0.088$ & $0.2$ & $516.9$ &  $3.99\times10^{-4}$ &  $2464.5$ \\
\hline
$n=1.0$                &                   &                          &                   &                          &                                              &                 &                       \\
\hline
$T_c=15$ MK & $0.661$ & $0.105$ & $0.775$ & $0.165$ & $113.9$ &  $0.67$ &  $5313.9$ \\
$T_c=10$ MK   & $0.225$ & $0.128$ & $0.430$ & $0.181$ & $103.3$ &  $4.62\times10^{-2}$ &  $3663.4$ \\
$T_c=8$ MK    & $0.113$ & $0.145$ & $0.229$ & $0.192$ & $97.8$ &  $9.40\times10^{-3}$ &  $3366.4$ \\
$T_c=4$ MK    & $0.062$ & $0.045$ & $0.094$ & $0.146$ & $516.3$ &  $4.05\times10^{-4}$ &  $2392.8$ \\
\hline
$n=1.5$                &                   &                          &                   &                          &                                              &                 &                       \\
\hline
$T_c=15$ MK & $0.558$ & $0.150$ & $0.729$ & $0.222$ & $114.2$ &  $0.65$ &  $5445.9$ \\
$T_c=10$ MK   & $0.192$ & $0.197$ & $0.371$ & $0.260$ & $103.4$ &  $4.59\times10^{-2}$ &  $3934.8$ \\
$T_c=8$ MK    & $0.109$ & $0.235$ & $0.219$ & $0.289$ & $97.7$ &  $9.41\times10^{-3}$ &  $3443.2$ \\
$T_c=4$ MK    & $0.062$ & $0.124$ & $0.094$ & $0.307$ & $516.7$ &  $4.05\times10^{-4}$ &  $2397.7$ \\
\hline \hline
\end{tabular}
\end{table}

Figure~\ref{fig:2DSolutions_comparison} and table~\ref{table:2d_comparison} present how the baryonic and dark components respond to the presence of DM at fixed core temperature. The four columns of the figure correspond to progressively smaller values of $T_c$, while the rows compare the pure-baryonic case with a constant-density DS, and two polytropic DS models with $n=1$ and $n=3/2$. The dashed magenta circles indicate the size of the DS, which allows us to compare between the size of the DS and the baryonic star.

We note that for a given $T_c$, the presence of DM reduces the baryonic mass required to sustain equilibrium. This is because the DM distribution deepens the gravitational potential and provides additional gravitational compression, which lowers the baryonic mass needed to reach ignition conditions. The effect is strongest in the constant-density case, where the DS is the most massive and compact and is milder for the polytropic models. This ordering is reflected in table~\ref{table:2d_comparison}. For example, for $T_c=15~\rm MK$, the no-DS model requires $M_b=1.140\,M_\odot$, while the constant-density DS yields only $M_b=0.219\,M_\odot$, and the polytropic cases lie in between. Thus, the more influential the dark component is, the smaller the baryonic mass needed to reach the same central temperature. An additional feature of the constant-density DS model at $T_c=15$ MK is the internal energy transport of the baryonic component. We note that for the core, it becomes convective, the intermediate region remains radiative, and the outer envelope turns convective again. This behavior is characteristic of more massive and more luminous main-sequence stars. In this case, this occurs because the constant-density DS produces a very compact and strongly bound configuration. The baryonic component must therefore support a larger energy flux through a smaller volume, leading to a greater central temperature gradient and, consequently, to a convective core. The appearance of this three-zone structure is another sign that an introduction of a substantial density of DM into the stellar system leads the baryonic configuration to behave more like a high-mass main-sequence star, despite having a much smaller baryonic mass than the corresponding no-DS solution.

In the case of the polytropes, the $n=1$ model is stiffer than the $n=3/2$ model, and this difference is reflected in the global properties of the DS. The softer $n=3/2$ configuration is generally more extended and more massive than the $n=1$ configuration, as seen from the values of $R_{\rm DS}$ and $M_{\rm DS}$ in table~\ref{table:2d_comparison}  and from the magenta circles in fig.~\ref{fig:2DSolutions_comparison}. Consequently, the $n=3/2$ model generally produces a somewhat stronger modification of the baryonic structure. Nevertheless, the difference between the two polytropic models is small when contrasted with the constant-density case. This can be understood from their different dark-matter density profiles. All three models have the same central DS density, but the density decreases outward in the polytropic configurations, whereas it remains fixed in the constant-density model up to $R_{\rm DS}=0.2R_\odot$. The constant-density configuration therefore contains a larger enclosed dark mass over much of the baryonic region and produces a stronger additional gravitational field, leading to a larger modification of the baryonic structure.

The geometric evolution of the configurations is also important. For the constant-density DS, $R_{\rm DS}$ is fixed at $0.2\,R_\odot$, while the baryonic radius decreases as $T_c$ decreases. At high $T_c$, the baryonic envelope extends beyond the dark core, while at low $T_c$ the baryonic component becomes fully embedded inside the DS. At this point, as seen in the upper left panel of fig.~\ref{fig:Comparisons}, all smaller core temperature solutions are independent of the precise size of the DS. For the polytropic models, by contrast, $R_{\rm DS}$ is not fixed but determined self-consistently. The DS radius and mass vary with $T_c$: as $T_c$ decreases, the baryonic core density does as well, so the baryonic perturbation to the DS is diminished. In turn, the polytropic DSs grow in both size and mass, tending to their respective unperturbed, DM-only configurations. However, for $T_c=4~\rm MK$, the baryonic density rises sharply, as is expected from the transition from gas-domination to degeneracy-domination in the baryonic EoS. Consequently, the DSs become much lighter and smaller. We note that this effect is most noticeable for the $n=1$ polytrope. 

The luminosity and effective temperature follow the same overall pattern, but here they are interpreted at fixed $T_c$ rather than at fixed baryonic mass. At all core temperatures, the constant-density model is less luminous than the corresponding no-DS case, yet it has a much smaller radius and therefore a higher $T_{\rm eff}$, as previously discussed in fig.~\ref{fig:HRDiagramDS}. In the case of the polytropic models, the situation is more complex because the DM distribution is itself modified by the presence of the baryons. As a result, the effective temperature is set by a non-linear interplay between the baryonic radius and the luminosity. This is precisely the same competition highlighted in fig.~\ref{fig:HRDiagramDS}. Since $L=4\pi R_b^2 \sigma T_{\rm eff}^4$, an increase in luminosity tends to raise $T_{\rm eff}$, whereas an increase in the baryonic radius tends to lower it. In the low-mass regime, where degeneracy pressure becomes more important and the baryonic equation of state is less sensitive to temperature, this competition is weak and the shift in $T_{\rm eff}$ remains small. At larger baryonic masses, where gas pressure dominates more strongly, the DM-induced thermal expansion of the baryonic component becomes more pronounced. In that case, the increase in radius can outweigh the luminosity enhancement, leading in some cases to a lower $T_{\rm eff}$ than in the no-DS case. 

At low $T_c$, the baryonic quantities become nearly identical across the polytropic models because the DSs are larger than the baryonic component and their inner density profiles are nearly flat over the region occupied by the baryons. In this regime, the baryonic star effectively evolves inside a background dark component whose local structure varies only weakly with the specific polytropic index. As a result, the baryons do not feel the detailed behavior of the DS near its outer regions, and the surface properties of the baryonic configuration become almost insensitive to the precise polytropic type. This is why $M_b$, $R_b$, $L$, and $T_{\rm eff}$ converge in both cases, even though the total DS mass and radius may still differ substantially.

In fig.~\ref{fig:DensityProfiles}, we show the baryonic density and temperature profiles obtained for two cases with fixed total baryonic mass in the presence of a constant-density DS. In the upper panels the DS density is fixed while its radius is varied, while in the lower panels the DS radius is kept constant, but its density is changed. These profiles make explicit the change in the baryonic density and temperature distributions, already suggested by the mass-radius and HR diagrams.

A first feature we note from the right panels is that, as the amount of DM increases, both the core and surface temperature of the baryons increase. This behavior can be explained by the virial theorem: adding DM to the baryonic star will deepen the gravitational potential. The virial theorem dictates that the increase of the absolute value of the gravitational potential is accompanied by a corresponding increase in kinetic energy and thus temperature. 
A second feature is the non-monotonic behavior of the baryonic radius. As shown in the insets, increasing the amount of DM, either by increasing $R_{\rm DS}$ at fixed $\rho_{\rm DS}$ (upper panels) or by increasing $\rho_{\rm DS}$ at fixed $R_{\rm DS}$ (lower panels), causes the baryonic radius $R_b$ to first increase, reach a maximum, and subsequently decrease. This behavior can be understood as a competition between the increase in gravity produced by the DS and the corresponding increase in the thermal support required to maintain hydrostatic equilibrium.
Adding DM increases the core temperature. Since the luminosity depends sensitively on the temperature, even a slight increase in the latter can cause a significant increase on the former. The larger energy output of the star leads to increased overall temperature in all layers, which in turn means larger gas pressure and therefore the star expands. The situation is somewhat similar to what happens in red giants. However, once sufficient DM mass has been added, gravity overturns the radius increase settling progressively into smaller values. This happens because at this stage the density of the admixed star is increasing as we will see and therefore it needs a smaller volume. This is in contrast to the red giant case because in red giants i) the core cannot be compressed further and ii) the core provides pressure. In the DM admixed case both i) and ii) are absent. 
A third feature of interest is the non-monotonic behavior of the baryonic core density as the DS radius increases (upper panels). We note that for $R_{\rm DS}\lesssim0.05R_\odot$ there is an increase in $\rho_{b,c}$, then for $0.05R_\odot\lesssim R_{\rm DS}\lesssim0.07R_\odot$ the core density decreases and afterwards for $0.07R_\odot\lesssim R_{\rm DS}$ this quantity increases until $R_{\rm DS}>R_b$. Initially  the core of the low mass star is convective and   behaves as a polytrope with $n=3/2$. This can be seen for example at the first star of the second row of fig.~\ref{fig:2DSolutions_comparison} which has a mass $\sim0.22M_{\odot}$. Therefore an increase in temperature is accompanied by a corresponding increase in density. Once we have added a sufficiently large quantity of DM to create a significant change in the luminosity, the star ceases to behave as a polytrope. The luminosity increase makes the temperature even higher. This creates substantial gas pressure since $P\sim \rho T$. However, there is still no need for a large pressure support because the added mass to the system is not yet a big fraction of the total mass. This can be seen by inspection of the baryonic pressure equation in eq.~(\ref{eq:stellar_structure_DS}). Note, as we mentioned earlier, that due to its sensitive dependence on the temperature, the luminosity is affected earlier than the stage where the mass affects significantly the hydrostatic equilibrium. This means that the correction in pressure does not have to be large and therefore the density has to reduce in order to counterbalance the larger temperature increase. This density reduction does not hold for long. As soon as the added DM mass becomes a significant fraction of the baryonic mass, the density goes up again. One reason for this is the fact that by now the star radius decreases and therefore we need to pack more mass into smaller volume. A second, and more profound reason, has to do with the fact that once both the mass and the luminosity contributions of the added DM become large, the Eddington approximation becomes valid making the star behave again as a polytrope and therefore an increase in temperature leads also to an increase in density (and pressure). The reason for the validity of the Eddington approximation is due to the fact that at this point $\kappa L/M$ becomes roughly constant.  

\section{Low-mass-star mass-luminosity relation}\label{sec:mass_luminosity}
In this section we provide mass-luminosity relations for the admixed stars. We first review the standard case of ordinary stars with no DM. 
\subsection{Ordinary Main Sequence Star}
To obtain the mass-luminosity relation for a purely convective, low-mass star, we consider how the luminosity scales in the interior and in the photosphere.
\paragraph{Interior scalings.} The core density and pressure scale as $\rho_c\sim M/R^3$ and $P_c\sim M^2/R^4$. By the ideal gas law, the core temperature scales as $T_c\sim P_c/\rho_c \sim M/R$. Then, by eq.~\eqref{eq:energy_prod}, we can write the total luminosity scaling as
\begin{equation}\label{eq:interior_lum}
    L\sim M \epsilon_c\sim M \rho_c^\lambda T_c^\nu\sim M^{\lambda+\nu+1}R^{-(3\lambda+\nu)},
\end{equation}
where we have written the energy generation rate at the core as $\epsilon_c\sim\rho_c^\lambda T_c^\nu$ for some indices $\lambda,\nu$.

\paragraph{Photospheric scalings.} Here we can make use of the optical depth condition of eq.~\eqref{eq:tau_boundary} to write $1\sim\tau\sim\kappa P_{\rm ph}/g$. By the ideal gas law, $P_{\rm ph}\sim \rho_{\rm ph} T_{\rm eff}$. We also replace $g\sim M/R^2$ and write the opacity as $\kappa\sim\rho_{\rm ph}^a T_{\rm eff}^{-b}$ for some indices $a,b$  to obtain the relation
\begin{equation}
    \rho_{\rm ph}\sim M^{\frac{1}{1+a}}R^{-\frac{2}{1+a}}T_{\rm eff}^{\frac{b-1}{1+a}}.
\end{equation}
For low masses, the star is purely convective, so the EoS is well approximated via the polytropic relation $P\sim\rho^{1+1/n}$ with $n=3/2$. Once again, for an ideal gas, we  obtain  $T\sim\rho^{1/n}$. This is valid throughout the whole star, so we may relate the core temperature with the effective surface temperature as $T_c\sim T_{\rm eff}(\rho_c/\rho_{\rm ph})^{1/n}$. Using the previous scaling relations, this yields the effective temperature scaling on mass and radius to be
\begin{equation}\label{eq:T_eff_std}
    T_{\rm eff}\sim M^\alpha R^\beta,\hspace{0.5cm}\alpha\equiv\frac{n(1+a)-a}{n(1+a)+1-b},\ \beta\equiv\frac{1+3a-n(1+a)}{n(1+a)+1-b}.
\end{equation}
Finally, at the surface the star satisfies the Stefan-Boltzmann law, so the luminosity can also be written as $L\sim R^2 T_{\rm eff}^4\sim M^{4\alpha}R^{2(1+2\beta)}$. Making use of this equation to write $R$ as a function of $L$ and $M$ and replacing into eq.~\eqref{eq:interior_lum}, we obtain the following mass-luminosity relation:
\begin{equation}
    L\sim M^\omega,\hspace{0.5cm}\omega={\frac{2[(\lambda+\nu+1)(1+2\beta)+2\alpha(3\lambda+\nu)]}{2(1+2\beta)+3\lambda+\nu}}.
\end{equation}
For a polytropic index of $n=3/2$, energy generation due to the pp-chain ($\lambda=1$,~$\nu\approx4.5$) and an opacity dominated by H$^{-}$ ($a\approx1$, $b\approx-7$), we obtain the typical scaling $L\sim M^{2.1}$.

\subsection{Addition of dark star}
\paragraph{Interior scalings.} In order to obtain an analytical result, we will model the DS as having constant density $\rho_{\rm DS}=\rm const$ and an extension larger than the baryonic component $R_{\rm DS}>R$. Then, we can write the total enclosed mass at the baryonic surface as the addition of the contribution of both components $M+m_{\rm DS}(R)=M+\frac{4\pi}{3}\rho_{\rm DS}R^3=M(1+\eta)$, where we have defined the quantity $\eta\equiv m_{\rm DS}(R)/M=\frac{4\pi}{3}\rho_{\rm DS}R^3/M$. By the baryonic hydrostatic equilibrium equation, the core pressure and temperature scalings are modified into $P_c\sim (1+\eta)M^2/R^4$ and $T_c\sim(1+\eta)M/R$. Hence, the total luminosity will now scale as
\begin{equation}\label{eq:L_DS_interior}
    L\sim M^{\lambda+\nu+1}R^{-(3\lambda+\nu)}(1+\eta)^{\nu}.
\end{equation}

\paragraph{Photospheric scalings.} With the DS addition, the gravitational acceleration now scales as $g\sim (1+\eta)M/R^2$, which gives the baryonic density at the photosphere
\begin{equation}
    \rho_{\rm ph}\sim M^{\frac{1}{1+a}}R^{-\frac{2}{1+a}}T_{\rm eff}^{\frac{b-1}{1+a}} (1+\eta)^{\frac{1}{1+a}}.
\end{equation}
Once again, using the relation $T\sim\rho^{1/n}$ and the previous scalings, we can write the effective temperature as
\begin{equation}
    T_{\rm eff}\sim M^\alpha R^\beta (1+\eta)^\gamma,\hspace{0.5cm} \gamma=\frac{n(1+a)+1}{n(1+a)+1-b},
\end{equation}
where $\alpha$ and $\beta$ are as in eq.~\eqref{eq:T_eff_std}. Once again, by the Stefan-Boltzmann law, we now have $L\sim M^{4\alpha}R^{2+4\beta}(1+\eta)^{4\gamma}$. Unlike the standard scenario, we cannot solve for $R$, as $\eta$ has an explicit dependence on both $M$ and $R$. In order to proceed, we equate the scaling of the luminosity obtained previously with that of eq.~\eqref{eq:L_DS_interior} and obtain the relation
\begin{equation}
    M^p R^{-q} (1+\eta)^{s}={\rm const},\hspace{0.5cm}p=\lambda+\nu+1-4\alpha,\ q=3\lambda+\nu+2+4\beta,\ s=\nu-4\gamma.
\end{equation}
Taking the natural logarithm and differentiating with respect to $\ln M$, we have
\begin{equation*}
    p-q\frac{d\ln R}{d\ln M}+s\frac{d\ln(1+\eta)}{d\ln M}=0.
\end{equation*}
We now treat $R$ as a function of the baryonic mass $M$, so directly from the definition of $\eta\sim R^3/M$, we have $d\eta/d\ln M=\eta[3\ (d\ln R/d\ln M)-1]$. With this result, we can rewrite the previous relation to obtain
\begin{equation}
    \frac{d\ln R}{d\ln M}=\frac{p-s\frac{\eta}{1+\eta}}{q-3s\frac{\eta}{1+\eta}}.
\end{equation}
Finally, to obtain the mass-luminosity relation, we note that if we have a scaling $L\sim M^\omega$, then $\omega=\frac{d\ln L}{d\ln M}$, so by taking the natural logarithm of $L\sim M^{4\alpha}R^{2+4\beta}(1+\eta)^{4\gamma}$ and differentiating with respect to $\ln M$, we obtain the result
\begin{equation}
    \omega=4\left(\alpha-\gamma\frac{\eta}{1+\eta}\right)+2\left(1+2\beta+6\gamma\frac{\eta}{1+\eta}\right)\frac{p-s\frac{\eta}{1+\eta}}{q-3s\frac{\eta}{1+\eta}}.
\end{equation}
For the same parameters ($\lambda=1$, $\nu\approx4.5$, $a\approx1$, $b\approx-7$) we can calculate the difference from the no DS result $\Delta\omega=\omega_{\eta}-\omega_{\eta=0}=77736\eta/(217(217+16\eta))$, which is explicitly positive, leading to a faster fall of the luminosity as baryonic mass is decreased, as we observe from the top left panel of fig.~\ref{fig:DSRelations}.

\section{Main sequence lifetime}
For a star with solar composition, we can approximate the lifetime as
\begin{equation}
    t_{\rm MS}\approx10^{10}~\mathrm{yr}\left(\frac{M_b}{M_\odot}\right)\left(\frac{L}{L_\odot}\right)^{-1},
\end{equation}
where $M_b$ is the baryonic mass of the star. For example, for the fixed masses considered in fig.~\ref{fig:DensityProfiles}, we have $t_{\rm MS}\approx338.9~\rm Gyr$ for $M_b=0.215M_\odot$ and $t_{\rm MS}\approx678.7~\rm Gyr$ for $M_b=0.119M_\odot$, for the standard (no DM) scenario.  On the other hand, the respective lifetimes when DM is introduced are provided in table~\ref{table:lifetimes}.

\begin{table}[h!]
\caption{Lifetime of baryonic stars when DM is included. We consider the same cases as in fig.~\ref{fig:DensityProfiles}, with the left columns corresponding to $M_b=0.215M_\odot$ and different DS radii, but fixed DS density $\rho_{\rm DS}=100~\mathrm{g}~\mathrm{cm}^{-3}$ and the right columns representing $M_b=0.119M_\odot$ and different DS density, but fixed DS radius $R_{\rm DS}>R_b$.}
\label{table:lifetimes}
\centering
\begin{tabular}{cc|cc}
\hline
\multicolumn{2}{c}{$\rho_{\rm DS}=100~\mathrm{g}~\mathrm{cm}^{-3}$} & \multicolumn{2}{|c}{$R_{\rm DS}>R_b$} \\
\hline
$R_{\rm DS}$ {[}$R_\odot${]} & $t_{\rm MS}$ {[}Gyr{]} & $\rho_{\rm DS}$ {[}g cm$^{-3}${]} & $t_{\rm MS}$ {[}Gyr{]} \\
\hline
$0.01$ & $338.6$ & $1$   & $661.3$     \\
$0.02$ & $332.8$ & $10$  & $551.7$ \\
$0.03$ & $316.8$ & $20$  & $440.8$ \\
$0.04$ & $290.8$ & $30$  & $337.7$ \\
$0.05$ & $255.9$ & $40$  & $236.4$ \\
$0.06$ & $216.1$ & $50$  & $155.6$ \\
$0.07$ & $177.1$ & $60$  & $109.1$ \\
$0.08$ & $140.2$ & $70$  & $69.6$  \\
$0.09$ & $110.0$ & $80$  & $46.0$  \\
$0.1$  & $80.9$  & $90$  & $33.6$  \\
$0.2$  & $4.3$   & $100$ & $24.6$  \\
$>R_b$ & $4.0$   & $200$ & $3.3$   \\
\end{tabular}
\end{table}
\noindent The lifetime decreases systematically as the DM content increases. For $M_b=0.215M_\odot$, increasing the DS radius from $0.01R_\odot$ to $0.2R_\odot$ reduces the estimated lifetime from $338.6$ to $4.3$ Gyr, corresponding to a reduction by a factor of about $80$. For the $M_b=0.119M_\odot$ sequence, increasing the DS density from $1$ to $200~\mathrm{g~cm^{-3}}$ reduces the lifetime from $661.3$ to $3.3$ Gyr, corresponding to a reduction by a factor of about $200$. Therefore, the reduction in lifetime closely tracks the increase in luminosity induced by the additional gravitational potential.

For  DS configurations with little DM, the effect is small. For example, at $M_b=0.215M_\odot$, a DS with $R_{\rm DS}=0.01R_\odot$ gives essentially the same lifetime as the standard model. Similarly, for $M_b=0.119M_\odot$, a DS with density $1~\mathrm{g~cm^{-3}}$ changes the lifetime only from $678.7$ to $661.3$ Gyr. However, as the DS contribution to the gravitational potential becomes stronger, the required central temperature increases, leading to a rapid increase in the nuclear energy generation rate and luminosity and hence to a much shorter lifetime.

The lifetimes obtained here should be interpreted as estimates based on the equilibrium stellar models rather than as results of a full stellar-evolution calculation. In particular, the expression above assumes a fixed fuel fraction and uses the luminosity of each equilibrium configuration to estimate the corresponding fuel-consumption timescale. A time-dependent treatment would additionally account for the evolution of the hydrogen abundance, changes in the stellar structure, and the resulting variation of the luminosity over the main-sequence phase. Nevertheless, the large differences between the standard and DS configurations show that the enhanced luminosity can substantially shorten the characteristic nuclear-burning timescale of low-mass stars. In fact, we see that the no-DS lifetimes exceed the age of the Universe by factors of approximately $25$--$50$ for the two masses considered here, while the strongest DS configurations have characteristic lifetimes of only a few Gyr. This shows the possibility that low-mass stars embedded in sufficiently dense DM environments could evolve significantly faster than their standard counterparts. In such environments, configurations that would remain on the main sequence for hundreds of Gyr in the absence of DM could instead undergo substantial hydrogen depletion on cosmologically relevant timescales.

\section{Redshift due to dark star}
An interesting consequence of the large DS configurations obtained in fig.~\ref{fig:DSRelations} for the DM EoS given by eq.~\eqref{eq:DM_EoS} is that their enormous masses and radii leave only a limited imprint on the internal baryonic structure. This behavior is evident in the mass-luminosity and mass-radius relations, where the solutions obtained for $m_\chi=2~\mathrm{MeV}/c^2$ and $m_\chi=5~\mathrm{MeV}/c^2$ are indistinguishable from the constant-density case. The reason is that the baryonic component occupies only a small fraction of the total DS volume. Since the central regions of these extended DSs have nearly constant density, the baryons probe only the local DM density surrounding them and are therefore largely insensitive to the detailed structure of the outer dark envelope. As a result, one would naively expect that standard stellar observables such as luminosity, effective temperature, and baryonic radius are unable to distinguish  DM distributions that share similar central densities but  substantially different total mass and radius. 
However, this is not necessarily the case once the effect of  gravitational redshift is taken into account. Although the baryons respond primarily to the local DM density, photons emitted from the stellar surface must escape through the full gravitational potential generated by the DS. Consequently, the observed redshift retains information about the total dark mass enclosed by the photon trajectory, even when the stellar structure itself does not. Two stars with nearly identical luminosities, effective temperatures, and radii may therefore exhibit different gravitational redshifts if they are embedded in DSs of different masses. In principle, spectroscopic measurements could thus provide a way of probing the total amount of DM associated with a DS.

This possibility is particularly relevant for the very extended configurations obtained for $m_\chi=2~\mathrm{MeV}/c^2$ and $m_\chi=5~\mathrm{MeV}/c^2$. While their baryonic stellar observables are the same as those of the constant-density model, their much larger total masses lead to a deeper gravitational potential. Consequently, gravitational redshift might affect significantly the emitted spectrum and luminosity.  In this section, we estimate the importance of this effect.

A photon emitted from the surface of a star is observed with a slightly smaller frequency due to gravitational redshift. In the weak field limit, we have the approximate equation
\begin{equation}\label{eq:redshift_std}
    \frac{\nu_{\rm obs}}{\nu_{\rm em}}\equiv(1+z)^{-1}=1+\frac{\Phi(R)}{c^2}=1-\frac{GM}{Rc^2}.
\end{equation}
In addition, the redshift will modify the observed spectrum $B_{\rm obs}(\nu_{\rm obs},T_{\rm obs})=B_{\rm obs}\left(\nu_{\rm obs},\frac{T_{\rm em}}{(1+z)}\right)$, as well as the observed luminosity $L_{\rm obs}=L_{\rm em}/(1+z)^2$.

\begin{figure}[h!]
	\centering
	\includegraphics[width=.68\textwidth]{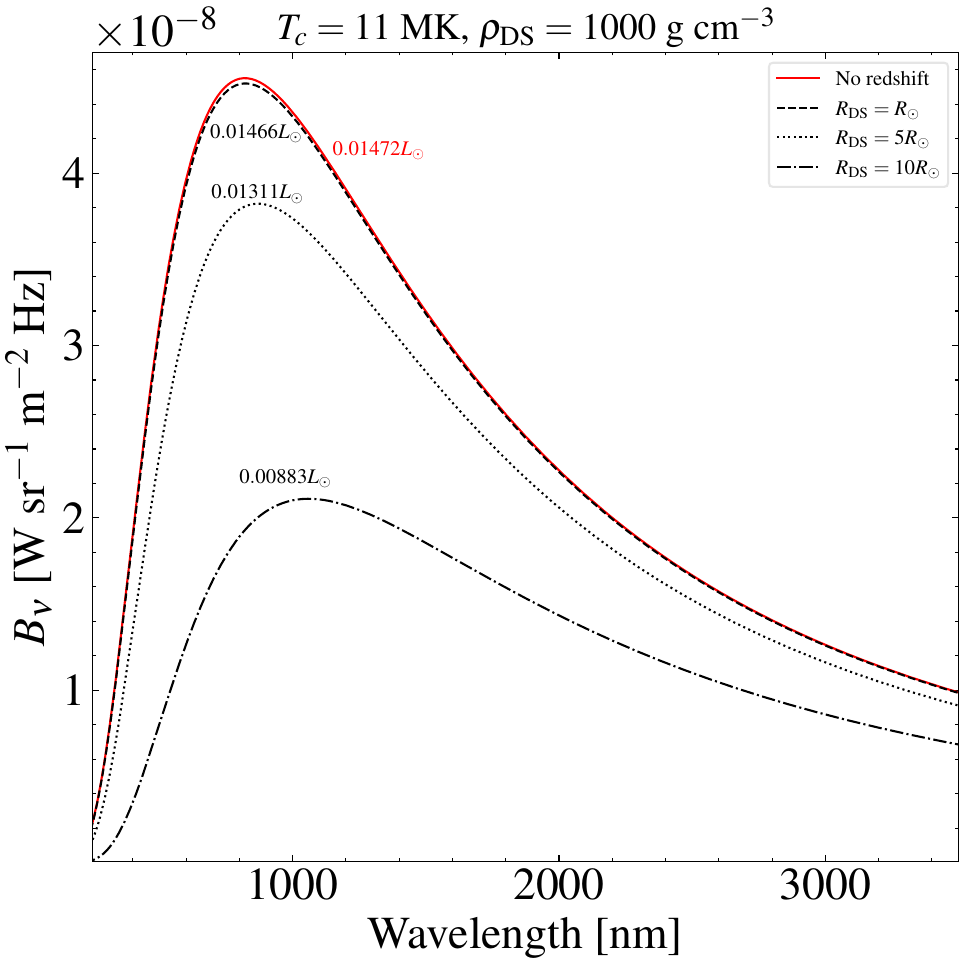}
	\caption{Change in the star's spectrum by extending the DM configuration. The DS density is fixed to $1000$ g cm$^{-3}$ and the core temperature is $11$ MK. The total baryonic mass is $0.014M_\odot$ and the baryonic radius is $0.084R_\odot$.}
	\label{fig:SpectrumModification}
\end{figure}

If we have the further contribution due to a DS, the potential must be modified to take this into account. Let the emission point be the baryonic star's radius $R_b$, then we have
\begin{equation}
    \Phi(R_b)=-\frac{GM_{<R_b}}{R_b}-4\pi G\int\displaylimits_{R_b}^{R_{\rm DS}}dr^\prime\,r^\prime\rho_{\rm DS}(r^\prime),
\end{equation}
where we define $M_{<R_b}\equiv \int\displaylimits_{0}^{R_b}dr^\prime\,4\pi {r^\prime}^2(\rho_{b}(r^\prime)+\rho_{\rm DS}(r^\prime))$ as the total enclosed mass at $r=R_b$. In the notation of section~\ref{sec:mass_luminosity}, we can write this term as $M_{<R_b}=M(1+\eta)$. For a constant density DS, we can analytically solve this potential and obtain
\begin{equation}
    \Phi(R_b)=
    \begin{cases}
        -\frac{GM}{R_b}-\frac{4\pi G \rho_{\rm DS} R_{\rm DS}^2}{3}\left(\frac{R_{\rm DS}}{R_b}\right),\hspace{2.3cm}\text{if }R_{\rm DS}<R_b,\cr
        -\frac{GM}{R_b}-2\pi G\rho_{\rm DS}R_{\rm DS}^2\left[1-\frac{1}{3}\left(\frac{R_b}{R_{\rm DS}}\right)^2\right],\hspace{0.5cm}\text{if }R_b<R_{\rm DS}.
    \end{cases}
\end{equation}
The addition of the DS will further reduce the frequency when compared with eq.~\eqref{eq:redshift_std}.

 In fig.~\ref{fig:SpectrumModification} we show some examples of how this effect can modify the observed stellar spectrum. We fix the DS density to $1000~\mathrm{g~cm^{-3}}$ and consider configurations with $R_{\rm DS}>R_b$. As we mentioned earlier, changing $R_{\rm DS}$ while keeping the density fixed has no effect on the internal baryonic structure when the DS  extends more than the baryonic star. However, the gravitational redshift depends on the potential generated by the entire DS. In the constant-density case, the contribution from the outer dark envelope scales approximately as $\Phi_{\rm DS}\propto \rho_{\rm DS}R_{\rm DS}^2$, so the observed signal can change substantially even when the baryonic mass, radius, and emitted spectrum remain  unchanged.

For the examples shown in the figure, increasing the DS radius from $R_{\rm DS}=5R_\odot$ to $R_{\rm DS}=10R_\odot$ produces a substantial increase in the gravitational redshift and reduces the observed luminosity by approximately $70\%$. At the same time, the characteristic temperature of the observed spectrum is shifted according to $T_{\rm obs}=T_{\rm em}/(1+z)$. Therefore, the effect of the DS not only reduces the observed luminosity, but it simultaneously shifts the spectrum toward lower observed energies and reduces the photon arrival rate. 

In principle one could distinguish different DS envelopes with the same central density. Independent measurements of luminosity, spectrum and the baryonic radius are sufficient to distinguish the DM envelopes. At the emission surface of the baryonic star the luminosity is related to the surface temperature and radius of the baryonic star through the relation $L_{\rm em}=4\pi R_b^2 \sigma T_{\rm em}^4$. If the effect of gravitational redshift is important and given the corresponding scalings of the luminosity and temperature with the redshift, setting $L_{\rm obs}=4\pi R_{b'}^{2}\sigma T_{\rm obs}^4$ will deduce a value for an ``effective" radius $R_{b'}$ that will be related to the true radius $R_b$ through the relation $R_{b'}=R_b(1+z)$. Therefore if the true baryonic radius can be measured, an indirect measurement of the redshift can be made. Although this will not be able to give us the exact density profile of the DS, it can provide an estimate of how massive and extended the DM configuration is.

\section{Conclusions}

In this work, we have studied the structure and observable properties of low-mass baryonic stars embedded in self-gravitating DM configurations. We have treated the baryonic and dark components as two gravitating fluids and assumed that the dark sector is non-luminous, interacting with the baryonic component only through gravity. Using the relevant hydrostatic equilibrium and energy transport equations, a realistic baryonic equation of state, opacity tables, and proton-proton-chain energy generation, we have investigated both constant-density and polytropic DM configurations, comparing the resulting stellar properties with those of ordinary DM-free stars.

The presence of DM substantially modifies the stellar structure.  At fixed baryonic mass, the enhanced gravitational field due to the presence of DM requires the baryonic component to develop greater thermal support, leading to higher interior temperatures and, consequently, enhanced nuclear energy generation. The resulting luminosity is therefore larger than in the corresponding ordinary DM-free stars, producing a modified mass-luminosity relation. The magnitude of this modification depends on both the density and the spatial extent of the DS. We also find that the baryonic radius does not vary monotonically with the DS properties, reflecting the competition between the stronger gravitational potential and the increased thermal support of the baryonic component. The enhanced luminosity due to the presence of DM has an additional important consequence: it lowers the minimum baryonic mass required for hydrogen fusion. Therefore depending on the amount of DM, the minimum mass for a baryonic star to sustain fusion can drop substantially below the ordinary  $\sim0.085M_\odot$ limit.

These structural changes have direct consequences for the stellar evolutionary timescale. Since the estimated main-sequence lifetime scales approximately as $t_{\rm MS}\propto M_b/L$ for fixed composition, the enhanced luminosity leads to substantially shorter lifetimes. For the benchmark configurations considered here, the characteristic lifetime can decrease from hundreds of Gyr in the standard case to only a few Gyr for more extreme DS configurations. These values should be interpreted as estimates based on equilibrium stellar models rather than as results of full time-dependent stellar-evolution calculations. Nevertheless, they demonstrate that a sufficiently strong dark gravitational potential can make the nuclear-burning timescale of otherwise extremely long-lived low-mass stars cosmologically relevant.

We have also shown that although the effect of a DM configuration does not change the internal stellar structure once it extends beyond the radius of the baryonic star, it can affect significantly both the observed luminosity and spectrum via gravitational redshift. Therefore, as we argued, an independent measurement of luminosity, spectrum and radius can reveal information regarding the mass and the extension of the DM configuration.

Our study demonstrates that stars that deviate from standard mass-radius and/or luminosity-mass relations can, in principle, be the signature of baryonic matter embedded in dense DM configurations. Masses inferred from binary systems are dynamical masses, so they probe the total mass of the star (baryonic plus DM). By contrast, in the standard stellar mass–radius relation, the relevant mass is the baryonic mass, assuming the radius being used is the baryonic radius. Deviations from ordinary stellar relations should therefore correlate with the local DM density profile. For example, stars with anomalous properties should be more common where the DM density is higher, such as near the Galactic center and in DM subhalos. Since DM is not confined to the Galactic plane, such atypical stars could also exist outside the plane. Furthermore the effect can be even more subtle. Due to the fact that under the influence of DM, the lifetime of main sequence stars can be reduced significantly, absence of relatively low mass stars in a cluster that by other means is known to be young can indirectly suggest the effect we have studied.  

There are several directions that open up for future work. One direction is to extend the present analysis to more massive stars, where among other changes, energy is mainly produced via the CNO cycle instead of the p-p chains. Moreover we intend to study supermassive Population III stars, for which the presence of a dark component could likewise modify the stellar structure and substantially alter their evolutionary timescales. Given that timescales are shortened significantly, it is interesting to investigate how the standard picture of stellar evolution is affected if, in cosmological redshifts relevant for galaxy formation, admixed stars like the ones studied here are present.
A further direction is the proper time-dependent treatment of the stellar evolution taking into account the change in the element composition due to fusion as time progresses.

\acknowledgments

The work of BBK is supported by IBS under the project code IBS-R018-D3.

\clearpage
\appendix
\section{pp‑chain energy production}\label{app:energy_injection}

The energy production rate is evaluated using the analytical fits to the reaction rates $N_{A}\langle\sigma v\rangle$ from the NACRE compilation~\cite{Angulo:1999zz} (with the exception of the $^7\mathrm{Be}+e^{-}\to\, ^{7}\mathrm{Li}+\nu_e$ electron capture rate, which was taken from Ref.~\cite{Caughlan:1987qf}). All quantities are expressed in SI units unless stated otherwise.

The total nuclear energy generation rate per unit mass $\varepsilon$, is obtained by summing the energy deposited in the stellar material by the individual reactions:
\begin{align}
\varepsilon &=
r_{\mathrm{pp}}(Q_{\mathrm{pp}}+Q_{\mathrm{Dp}})\cr
&+r_{^3{\rm He}^3{\rm He}}Q_{^3{\rm He}^3{\rm He}}
+r_{^3{\rm He}^4{\rm He}}Q_{^3{\rm He}^4{\rm He}}
+r_{^7{\rm Be}\,{\rm e}^-}Q_{^7{\rm Be}\,{\rm e}^-}
+r_{^7{\rm Li}\,{\rm p}}Q_{^7{\rm Li}\,{\rm p}}
+r_{^7{\rm Be}\,{\rm p}}Q_{^7{\rm Be}\,{\rm p}}.
\end{align}
Here, $r_i$ denotes the number of reactions of type $i$ occurring per unit mass and per unit time, while $Q_i$ denotes the energy deposited in the stellar material per reaction. The energy deposited per reaction is obtained from the nuclear $Q$‑values:
\begin{align}
&Q_{\mathrm{pp}} = (0.15+1.02)\, \mathrm{MeV},\cr
&Q_{\mathrm{Dp}} = 5.49\, \mathrm{MeV}, \cr
&Q_{^3{\rm He}^3{\rm He}} = 12.86\, \mathrm{MeV},\cr
&Q_{^3{\rm He}^4{\rm He}} = 1.59\, \mathrm{MeV},\cr
&Q_{^7{\rm Be}\,{\rm e}^-} = 0.05\, \mathrm{MeV}, \cr
&Q_{^7{\rm Li}\,{\rm p}} = 17.35\, \mathrm{MeV},\cr
&Q_{^7{\rm Be}\,{\rm p}} = (0.14+1.02+6.88+3.0)\, \mathrm{MeV}. 
\end{align}
These are not the total nuclear $Q_i$ values as the energy carried away by neutrinos is excluded. For the $p+p$ reaction, the term $0.15~\mathrm{MeV}$ accounts for the average energy retained from the reaction after neutrino energy loss, while $1.02~\mathrm{MeV}$ corresponds to positron annihilation with an electron. For the ${}^7\mathrm{Be}+p$ branch, the four terms correspond to the $0.14~\mathrm{MeV}$ $\gamma$-ray from ${}^7\mathrm{Be}(p,\gamma){}^8\mathrm{B}$, the $1.02~\mathrm{MeV}$ from positron annihilation, the $6.88~\mathrm{MeV}$ mean kinetic energy of the positron emitted in the subsequent ${}^8\mathrm{B}\to{}^8\mathrm{Be}+e^++\nu_e$ decay, and the $3.0~\mathrm{MeV}$ released as kinetic energy of the two $\alpha$-particles from ${}^8\mathrm{Be}\to2\,{}^4\mathrm{He}$. The $\sim7~\mathrm{MeV}$ carried away by the ${}^8\mathrm{B}$ neutrino is excluded from the deposited energy.
\subsection{Reaction rates $N_{A}\langle\sigma v\rangle$}
We consider the following rates in units ${\rm cm}^3~\mathrm{mol}^{-1}~ {\rm s}^{-1}$, where $N_{A}$ is Avogadro's number:
\allowdisplaybreaks
\begin{align}
&N_{A}\langle\sigma v\rangle_{\mathrm{pp}} = 4.08\times10^{-15}\,T_9^{-2/3}
\mathrm{e}^{-3.381\,T_9^{-1/3}}
\left(1+3.82\,T_9+1.51\,T_9^2+0.144\,T_9^3
-1.14\times10^{-2}\,T_9^4\right),\cr
&N_{A}\langle\sigma v\rangle_{^3{\rm He}^3{\rm He}} = 5.59\times10^{10}\,T_9^{-2/3}
\mathrm{e}^{-12.277\,T_9^{-1/3}}
\left(1-0.135\,T_9+2.54\times10^{-2}\,T_9^2
-1.29\times10^{-3}\,T_9^3\right),\cr
&N_{A}\langle\sigma v\rangle_{^3{\rm He}^4{\rm He}} = 5.46\times10^{6}\,T_9^{-2/3}
\mathrm{e}^{-12.827\,T_9^{-1/3}}
\Bigl(1-0.307\,T_9+8.81\times10^{-2}\,T_9^2
\nonumber\\
&\qquad\hspace{5.8cm}
-1.06\times10^{-2}\,T_9^3
+4.46\times10^{-4}\,T_9^4\Bigr),\cr
&N_{A}\langle\sigma v\rangle_{^7{\rm Be}\,{\rm e}^-} = 1.34\times10^{-10}\,T_9^{-1/2}
\left(1-0.537\,T_9^{1/3}+3.86\,T_9^{2/3}
+0.0027\,T_9^{-1}\mathrm{e}^{2.515\times10^{-3}\,T_9^{-1}}\right),\cr\cr\cr
&N_{A}\langle\sigma v\rangle_{^7{\rm Li}\,{\rm p}} = 7.2\times10^{8}\,T_9^{-2/3}
\mathrm{e}^{-8.473\,T_9^{-1/3}-(T_9/6.5)^2}
\Bigl(1+1.05\,T_9-0.653\,T_9^2+0.185\,T_9^3
\nonumber\\
&\qquad\hspace{7.2cm}
-2.12\times10^{-2}\,T_9^4
+9.3\times10^{-4}\,T_9^5\Bigr)
\nonumber\\
&\qquad\hspace{7.3cm}
+9.85\times10^{6}\,T_9^{0.576}
\mathrm{e}^{-10.415\,T_9^{-1}},\cr
&N_{A}\langle\sigma v\rangle_{^7{\rm Be}\,{\rm p}} = 2.61\times10^{5}\,T_9^{-2/3}
\mathrm{e}^{-10.264\,T_9^{-1/3}}
\Bigl(1-5.11\times10^{-2}\,T_9
+4.68\times10^{-2}\,T_9^2-6.6\times10^{-3}\,T_9^3
\nonumber\\
&\qquad\hspace{6.0cm}
+3.12\times10^{-4}\,T_9^4\Bigr)
+2.05\times10^{3}\,T_9^{-3/2}
\mathrm{e}^{-7.345\,T_9^{-1}}.
\end{align}
The temperature is expressed in units of $10^9\ \rm K$: $T_9 = T\times10^{-9}$. The conversion to particle rates $\lambda_i$ in units ${\rm m}^3~ {\rm s}^{-1}$  is performed as
\begin{equation}
\lambda_i = \frac{10^{-6}}{N_{\!A}}\,N_{\!A}\langle\sigma v\rangle_i,
\end{equation}
For the $^7$Be electron capture rate, the fit contains the term $\mathrm{exp}(2.515\times10^{-3}/T_9)$ which is not numerically reliable at very low temperatures. Therefore, for $T<10^6~\rm K$, the electron capture rate per $^7$Be nucleus $\Gamma_{^7{\rm Be}\,{\rm e}^-}\equiv \lambda_{^7{\rm Be}\,{\rm e}^-}n_e$ is capped at $1.57\times10^{-7}~\mathrm{s}^{-1}$.

\subsection{Number densities}
The local composition is specified by the hydrogen mass fraction $X$ and the total helium mass fraction $Y$, which includes both $^3\mathrm{He}$ and $^4\mathrm{He}$. The mass fractions of the intermediate nuclei $^7\mathrm{Li}$ and $^7\mathrm{Be}$ are denoted by $Z_{\mathrm{Li}}$ and $Z_{\mathrm{Be}}$, respectively. Using the atomic mass unit $m_{\mathrm{u}}=1.661\times10^{-27}\,\mathrm{kg}$, the number densities are
\begin{align}\label{eq:nuclear_densities}
n_{\mathrm{p}} = \frac{\rho X}{m_{\mathrm{u}}},\quad
n_{\mathrm{^4He}} &= \frac{\rho(Y-Y_3)}{4\,m_{\mathrm{u}}},\quad
n_{\mathrm{^3He}} = \frac{\rho Y_3}{3\,m_{\mathrm{u}}},\cr
n_{\mathrm{^7Be}} = \frac{\rho Z_{\mathrm{Be}}}{7\,m_{\mathrm{u}}},\quad
n_{\mathrm{^7Li}} &= \frac{\rho Z_{\mathrm{Li}}}{7\,m_{\mathrm{u}}},\quad
n_{\mathrm{e}} = n_{\mathrm{p}} + 2n_{\mathrm{^3He}} + 2n_{\mathrm{^4He}} + 4n_{\mathrm{^7Be}} + 3n_{\mathrm{^7Li}},
\end{align}
where the electron number density is obtained by imposing charge neutrality.

\subsection{Intermediate abundances in the steady-state regime}
The abundances of the intermediate nuclei are determined algebraically by imposing local steady state. For two-body reactions, the reaction rate per unit volume is
\begin{equation*}
\mathcal{R}_{ij} =
\frac{n_i n_j}{1+\delta_{ij}}\,\lambda_{ij},
\end{equation*}
where $\delta_{ij}=1$ for identical nuclei and is zero otherwise. The corresponding rate per unit mass is $r_{ij}=\mathcal{R}_{ij}/\rho$. The $^3\mathrm{He}$ abundance is found by equating its production rate to its destruction rate,
\begin{equation*}
\mathcal{R}_{\mathrm{pp}}
=
2\mathcal{R}_{^3\mathrm{He}^3\mathrm{He}}
+\mathcal{R}_{^3\mathrm{He}^4\mathrm{He}}.
\end{equation*}
Using the number densities defined above, this condition becomes
\begin{equation*}
\frac{\lambda_{\mathrm{pp}}X^2}{2} = \frac{\lambda_{^3\mathrm{He}^3\mathrm{He}}}{9}Y_3^2 + \frac{\lambda_{^3\mathrm{He}^4\mathrm{He}}}{12} Y_3(Y-Y_3).
\end{equation*}
The physical solution is obtained from the positive root of this quadratic equation, subject to $0\leq Y_3\leq Y$. The $^4\mathrm{He}$ abundance is then determined by the remaining helium mass fraction, $Y-Y_3$. If no physical solution satisfies this constraint, the steady-state prescription is not applicable to the specified local conditions.

The $^7\mathrm{Be}$ abundance is determined by balancing its production through $^3\mathrm{He}+{}^4\mathrm{He}$ against its destruction through electron capture and proton capture:
\begin{equation*}
\mathcal{R}_{^3\mathrm{He}^4\mathrm{He}} = \mathcal{R}_{^7\mathrm{Be}\,e^-} + \mathcal{R}_{^7\mathrm{Be}\,p}.
\end{equation*}
Defining the electron-capture rate per $^7\mathrm{Be}$ nucleus as $\Gamma_{^7{\rm Be}\,{\rm e}^-}\equiv \lambda_{^7{\rm Be}\,{\rm e}^-}n_e$, this yields
\begin{equation*}
n_{\mathrm{^7Be}} = \frac{ \lambda_{^3\mathrm{He}^4\mathrm{He}} n_{\mathrm{^3He}}n_{\mathrm{^4He}} }{ \Gamma_{^7{\rm Be}\,{\rm e}^-}+\lambda_{^7\mathrm{Be}\,p}n_{\mathrm{p}} }.
\end{equation*}
The $^7\mathrm{Li}$ abundance is obtained by balancing production through electron capture against destruction by proton capture,
\begin{equation*}
\mathcal{R}_{^7\mathrm{Be}\,e^-} = \mathcal{R}_{^7\mathrm{Li}\,p},
\end{equation*}
which gives
\begin{equation*}
n_{\mathrm{^7Li}} = \frac{ \Gamma_{^7{\rm Be}\,{\rm e}^-}n_{\mathrm{^7Be}} }{\lambda_{^7\mathrm{Li}\,p}n_{\mathrm{p}} }.
\end{equation*}
The corresponding mass fractions are calculated from $Y_3=3m_{\mathrm{u}}n_{\mathrm{^3He}}/\rho$, $Z_{\mathrm{Be}}=7m_{\mathrm{u}}n_{\mathrm{^7Be}}/\rho$, and $Z_{\mathrm{Li}}=7m_{\mathrm{u}}n_{\mathrm{^7Li}}/\rho$.

Finally, because the electron number density depends on the $^7\mathrm{Be}$ and $^7\mathrm{Li}$ abundances (see eq.~\eqref{eq:nuclear_densities}), while $\Gamma_{^7{\rm Be}\,{\rm e}^-}$ in turn depends on $n_e$, the system is closed by fixed-point iteration. Starting from the fully ionised H and He value $n_e^{(0)}=n_p+2n_{^3\rm He}+2n_{^4\rm He}$, the abundances $n_{^7\rm Be}$ and $n_{^7\rm Li}$ are evaluated from the balance equations above, $n_e$ is updated from the resulting charge density, and the procedure is repeated. Three iterations are sufficient to reach convergence.

\subsection{Per‑mass reaction rates and steady‑state adjustments}
Once the intermediate abundances have been determined from the steady-state conditions in the preceding subsection, the per‑mass reaction rates (in units kg$^{-1}$ s$^{-1}$) are evaluated as
\begin{align}
r_{\mathrm{pp}} &= \frac{n_{\mathrm{p}}^2}{2\rho}\lambda_{\mathrm{pp}},\quad
r_{^3{\rm He}^3{\rm He}} = \frac{n_{\mathrm{^3He}}^2}{2\rho}\lambda_{^3{\rm He}^3{\rm He}},\quad
r_{^3{\rm He}^4{\rm He}} = \frac{n_{\mathrm{^3He}}\,n_{\mathrm{^4He}}}{\rho}\lambda_{^3{\rm He}^4{\rm He}},\\
r_{^7{\rm Be}\,{\rm e}^-} &= \frac{n_{\mathrm{^7Be}}\,\Gamma_{^7{\rm Be}\,{\rm e}^-}}{\rho},\quad
r_{^7{\rm Li}\,{\rm p}} = \frac{n_{\mathrm{^7Li}}\,n_{\mathrm{p}}}{\rho}\lambda_{^7{\rm Li}\,{\rm p}},\quad
r_{^7{\rm Be}\,{\rm p}} = \frac{n_{\mathrm{^7Be}}\,n_{\mathrm{p}}}{\rho}\lambda_{^7{\rm Be}\,{\rm p}}.
\end{align}
Because $Y_3$, $n_{\mathrm{^7Be}}$ and $n_{\mathrm{^7Li}}$ are obtained by imposing the balance equations of the previous section, these rates automatically satisfy
\begin{equation*}
    r_{\mathrm{pp}}=2r_{^3{\rm He}^3{\rm He}}+r_{^3{\rm He}^4{\rm He}},\hspace{0.5cm}r_{^3{\rm He}^4{\rm He}}=r_{^7{\rm Be}\,{\rm e}^-}+r_{^7{\rm Be}\,{\rm p}},\hspace{0.5cm}r_{^7{\rm Be}\,{\rm e}^-}=r_{^7{\rm Li}\,{\rm p}}
\end{equation*}
These rates are then inserted into the expression for $\varepsilon$ given in the main text by eq.~\eqref{eq:energy_generation}.

\bibliographystyle{JHEP} 
\bibliography{References} 
      
\end{document}